\documentclass[aps,showpacs]{revtex4}
\usepackage{amsfonts}
\usepackage{amsmath}
\usepackage{amssymb}
\usepackage{graphicx}
\usepackage{fancyheadings}
\usepackage{times,xspace,color}
\usepackage{bm}

\def\fn{\mathfrak{n}}

\def\fS{{\sf{S}}}

\def\one{{\mathchoice {\rm 1\mskip-4mu l} {\rm 1\mskip-4mu l} {\rm
1\mskip-4.5mu l} {\rm 1\mskip-5mu l}}}
\def\bbbc{{\mathchoice {\setbox0=\hbox{$\displaystyle\rm C$}\hbox{\hbox
to0pt{\kern0.4\wd0\vrule height0.9\ht0\hss}\box0}}
{\setbox0=\hbox{$\textstyle\rm C$}\hbox{\hbox
to0pt{\kern0.4\wd0\vrule height0.9\ht0\hss}\box0}}
{\setbox0=\hbox{$\scriptstyle\rm C$}\hbox{\hbox
to0pt{\kern0.4\wd0\vrule height0.9\ht0\hss}\box0}}
{\setbox0=\hbox{$\scriptscriptstyle\rm C$}\hbox{\hbox
to0pt{\kern0.4\wd0\vrule height0.9\ht0\hss}\box0}}}}

\newcommand{\Integers}{\ensuremath{\mathbb{Z}}\xspace}

\newcommand{\ket}[1]{|{#1}\rangle}

\newcommand{\bi}{{\bf i}}
\newcommand{\bj}{{\bf j}}
\newcommand{\bk}{{\bf k}}
\newcommand{\bl}{{\bf l}}

\preprint{}
\begin{document}
\title{Exactly-Solvable Models Derived from a Generalized Gaudin Algebra}

\author{ G. Ortiz}
\email{ortiz@viking.lanl.gov}

\author{ R. Somma }
\email{somma@viking.lanl.gov} \affiliation{Los Alamos National
Laboratory, Los Alamos, New Mexico 87545, USA}

\author{ J. Dukelsky}
\email{dukelsky@iem.cfmac.csic.es} \affiliation{Instituto de
Estructura de la Materia, CSIC, Serrano 123, 28006 Madrid, Spain}

\author{S.~Rombouts}
\email{Stefan.Rombouts@rug.ac.be}
\affiliation{Ghent University, Department of Subatomic and Radiation Physics,
Proeftuinstraat 86, B-9000 Gent, Belgium.}

\date{Received \today }

\begin{abstract}
We introduce a generalized Gaudin Lie algebra and a complete set
of mutually commuting quantum invariants allowing the derivation
of several families of exactly solvable Hamiltonians. Different
Hamiltonians correspond to different representations of the
generators of the algebra. The derived exactly-solvable
generalized Gaudin models include the Bardeen-Cooper-Schrieffer,
Suhl-Matthias-Walker, the Lipkin-Meshkov-Glick, generalized Dicke,
the Nuclear Interacting Boson Model, a new exactly-solvable
Kondo-like impurity model, and many more that have not been
exploited in the physics literature yet.

\end{abstract}


\maketitle

\section{Introduction}
\label{sec1}

During last decade we have witnessed an enormous progress both in
low-temperature experimental techniques and in the design and better
characterization of novel materials and cold atomic systems. These
developments allow one to explore the quantum world in a more
fundamental way. In particular, since interactions between particle
constituents can lead to unexpected phenomena, one would like to
achieve sufficient degree of quantum control to take advantage of it.
Theoretical work on strongly coupled systems is helping in this regard.
For example, research on critical phenomena in quantum phase transitions,
where the Landau-Ginzburg paradigm of broken-symmetry phase transitions
does not apply, shows interesting scenarios. Such is the case of the
recently proposed deconfined quantum critical points where
fractionalized excitations may emerge at criticality with observable
consequences \cite{senthil}. A crucial theoretical bottleneck,
however, is the lack of exactly-solvable interacting many-body models,
since non-perturbative and non-linear phenomena play a relevant role.

The main goal of this paper is to introduce a generalization of the
Gaudin algebra \cite{Gau1}, which we name {\it generalized Gaudin
algebra} (GGA), whose quantum invariants can be exactly diagonalized
and may be related to Hamiltonian operators of exactly-solvable
problems of interacting constituents. By exactly-solvable model we mean
a model Hamiltonian whose entire spectral problem is reduced to an
algebraic one (i.e., it is explicitly diagonalized), a fact that  is
associated to the existence of a certain hidden symmetry in the model
under consideration.  There are larger classes of Hamiltonians
characterized by exact solvability of only certain part of their
spectra; these are called {\it quasi-exactly} solvable \cite{ushve},
and the $t$-$J_z$ chain model is an example \cite{tJz}. Clearly,
exactly-solvable models may be used as a starting point to construct
many other quasi-exactly solvable models.

As we will see, we have identified the main operator algebra underlying
the integrability and exact solvability of  many well-known models,
thus unifying their description in a single algebraic framework. Simply
diagonalizing the quantum invariants of the GGA is sufficient to solve
all those problems, which include the Bardeen-Cooper-Schrieffer (BCS)
\cite{bcs}, Suhl-Matthias-Walker (SMW) \cite{SMW}, Lipkin-Meshkov-Glick
(LMG) \cite{limegl}, generalized Dicke (GD) \cite{Dicke}, and many
others of interest in condensed-matter, molecular, atomic and nuclear
physics. The basic point is that all these various models, which form
the general class of $XYZ$ Gaudin models, can be derived using
different realizations of the generators of the GGA. For example, the
BCS model is obtained from the quantum invariants of the GGA after
representing their generators in terms of fermionic-pair realizations
of the generators of $\bigoplus_\bl su(2)$. A consequence of this
unification is that {\it new} exactly-solvable models can be realized
after a proper representation of the GGA. For instance, one can write
down exactly-solvable $SU(N)$ spin and mixed representation models,
such as spin-fermion, spin-boson or fermion-boson Hamiltonians.

We start by defining the GGA in section \ref{sec2}. We show how the
$XYZ$ Gaudin equation naturally emerges from the Jacobi identity for
the generators of the GGA. We also introduce the quantum invariants
that will serve as the generating functions for all conserved
quantities of the generalized (integrable) $XYZ$ Gaudin models.

In section \ref{sec3} the $XXZ$ Gaudin equation and the diagonalization
of the $XXZ$ Gaudin models are studied. We show a family of solutions of
the $XXZ$ Gaudin equation, which includes the well-known rational,
trigonometric, hyperbolic \cite{2D, Ami} and the new solution found by
Richardson \cite{Richx} as especial limits. In particular, we show that
the latter can be considered as a reparametrization of the other three.
The main use of these solutions is to design exactly-solvable
Hamiltonians with a large set of free parameters, thus providing
additional freedom to tune interactions.

In section \ref{sec4} we consider two possible realizations of the GGA
in terms of the generators of $\bigoplus_\bl su(2)$ and $\bigoplus_\bl
su(1,1)$ which allow us to construct (given the analytic properties of
the solutions of the Gaudin equation) general Gaudin model Hamiltonians
that will be exploited in the rest of the paper. Clearly, from the
oscillator realizations of $\bigoplus_\bl su(2)$ and $\bigoplus_\bl
su(1,1)$ in terms of canonical fermions and bosons, one can build
several interesting many-body Hamiltonians, including the BCS, SMW,
LMG, and GD. But one is not limited to these oscillators realizations.
Indeed, one can use, for instance,  $SU(N)$ or hard-core particles
realizations to construct new exactly-solvable Hamiltonians
\cite{reviewbo}.

Sections \ref{sec5} and \ref{sec6} present applications of the
algebraic framework to various well-known models. They correspond to
different realizations of the algebras $su(2)$ or $su(1,1)$ in terms of
canonical fermions or bosons. We start section \ref{sec5} by solving
the BCS pairing models in an arbitrary basis and then focus on the
analysis of the BCS Hamiltonian in momentum space. We study,
particularly, multiband pairing Hamiltonians such as the SMW model
which is of relevance for the description of two-gap superconductivity
in MgB$_2$. In section \ref{sec6}, we analyze bosonic pairing models of
interest in cold atom physics. In particular, section \ref{sec6c}
concentrates on exactly-solvable two-level boson Hamiltonians, which
include the nuclear Interacting Boson Model (IBM) \cite{IBM}, the two
Josephson-coupled Bose-Einstein condensates (BECs) model \cite{Jop},
and the LMG model. Most importantly, we show that the LMG model, used
for decades to study phase transitions in finite nuclei, is exactly
solvable.

One would like to understand what are the general differences between
mean-field approximations to the $XYZ$ Gaudin models, and their exact
solution. As we will see, for finite systems the distinction is evident
and the character of the solutions of the two approaches differs
substantially. However, does the difference persist in the
thermodynamic limit, i.e. the infinite ($N \rightarrow \infty$)
system-size limit? Expectation values of certain observables  (e.g.,
the BCS gap equation or the occupation numbers) will be identical, but
other observables may pick up the differences. A question that
naturally arises concerns the critical behavior of the $XYZ$ Gaudin
models. One would like to know, for example, what are their quantum
critical exponents. It turns out that for certain Gaudin models (e.g.,
the LMG and SMW model of Eq. (\ref{H2Bn})) the critical behavior is
mean-field \cite{rolo}. This is very simple to prove by applying tools
from Lie algebras and Catastrophe theory, as developed by R. Gilmore
\cite{cata}. The general analysis is beyond the scope of the present
paper and will be presented in a separate publication.
Here, however, we will only analyze the quantum phase diagram of the
BCS model as a function of the interaction strength and show that the
transition between the superconducting and Fermi-liquid phases is
Kosterlitz-Thouless-like, independently of the space dimensionality of
the lattice.

Mixing realizations and representations of the generators of the
Gaudin algebra lead to new exactly-solvable models.  In section
\ref{sec7}  we illustrate these ideas by solving three types of
many-body models: the GD, an exactly-solvable Kondo-like, and a
spin-boson models. In this way, one finds the formal algebraic
connection between these different physical phenomena and BCS
superconductivity. Section \ref{sec8} deals with differential
operator realizations of the Gaudin generators leading to
quasi-exactly solvable problems in the continuum. Finally, we show
in the Appendix that the weak-coupling limit solutions of the
generalized $XXZ$ Gaudin models are given by the roots of Laguerre
polynomials.

%

\section{Generalized Gaudin algebras}
\label{sec2}

\subsection{Commutation relations}

Let us introduce the GGA as the set of operators
$\{\fS^\kappa_m \equiv \fS^\kappa(E_m)\}$, with $\kappa=x,y,z$,
satisfying the commutation relations (for $E_m \neq E_\ell$)
\begin{eqnarray}
\begin{cases}
[\fS^\kappa_m,\fS^\kappa_\ell]=0 \ ,  \\
[\fS^x_m,\fS^y_\ell]= i (Y_{m\ell} \ \fS^z_m- X_{m\ell} \ \fS^z_\ell) \ ,
 \\
[\fS^y_m,\fS^z_\ell]= i (Z_{m\ell} \ \fS^x_m- Y_{m\ell} \ \fS^x_\ell) \ ,
 \\
[\fS^z_m,\fS^x_\ell]= i (X_{m\ell} \ \fS^y_m- Z_{m\ell} \ \fS^y_\ell) \ ,
\end{cases}
\label{ggaudin}
\end{eqnarray}
where $X_{m\ell}=X(E_m,E_\ell)$, $Y_{m\ell}=Y(E_m,E_\ell)$, and
$Z_{m\ell}=Z(E_m,E_\ell)$ are antisymmetric (i.e. $W(x,y)=-W(y,x)$)
complex functions of two arbitrary complex variables $E_m,
E_\ell$ labeled by positive integers $m$ and $\ell$, respectively.
Equivalently, in terms of the  $\kappa=+,-,z$ basis (and for $E_m \neq
E_\ell$)
\begin{eqnarray}
\begin{cases}
[\fS^\pm_m,\fS^\pm_\ell]=\pm 2 \ V^-_{m\ell} \ (\fS^z_m+\fS^z_\ell) \ ,
\\
[\fS^-_m,\fS^+_\ell]= - 2 \ V^+_{m\ell} \ (\fS^z_m-\fS^z_\ell) \ ,
 \\
[\fS^z_m,\fS^\pm_\ell]= \pm ( V^+_{m\ell} \ \fS^\pm_m- Z_{m\ell} \
\fS^\pm_\ell - V^-_{m\ell} \ \fS^\mp_m)\ ,
\end{cases}
\label{ggaudin2}
\end{eqnarray}
where $\fS^\pm_m=\fS^x_m \pm i\fS^y_m$, and
$V^\pm_{m\ell}=(X_{m\ell}\pm Y_{m\ell})/2$. (Notice that $\fS^{+(-)}_m$
and $\fS^{+(-)}_\ell$ are non-commuting operators, unless
$X_{m\ell}=Y_{m\ell}$.)

The complex functions $X_{m \ell},Y_{m \ell}$, and $Z_{m \ell}$ are
taken to have the limiting behavior
\begin{eqnarray}
\lim_{\varepsilon \rightarrow 0} \ \varepsilon X(x,x+\varepsilon) =
{\sf f}(x) \ , \
\lim_{\varepsilon \rightarrow 0} \ \varepsilon Y(x,x+\varepsilon) =
{\sf g}(x) \ , \
\lim_{\varepsilon \rightarrow 0} \ \varepsilon Z(x,x+\varepsilon) =
{\sf h}(x) \ ,
\label{limiting}
\end{eqnarray}
where ${\sf f}(x)$, ${\sf g}(x)$, and  ${\sf h}(x)$ are non-singular
functions. Indeed, $X, Y$, and $Z$ are complex meromorphic functions
having poles of order one. In particular, when  ${\sf f}(x)={\sf
g}(x)={\sf h}(x)$ the above commutation relations,  Eqs.
(\ref{ggaudin}), can be analytically continued to the case $m=\ell$
(i.e., $E_m \rightarrow E_\ell$). For example,
\begin{equation}
[\fS^x_m,\fS^y_m]= \lim_{\varepsilon \rightarrow 0}  i (Y(E_m,E_m +
\varepsilon) \ \fS^z(E_m)- X(E_m,E_m+\varepsilon) \ \fS^z(E_m
+\varepsilon))=  -i \ {\sf f}(E_m) \frac{\partial \fS^z_m}{ \partial E_m}
\ .
\nonumber
\end{equation}
Then,
\begin{eqnarray}
\begin{cases}
[\fS^\kappa_m,\fS^\kappa_m]=0 \ , \\
[\fS^x_m,\fS^y_m]= -i \ {\sf f}(E_m) \frac{\partial \fS^z_m}{ \partial E_m} \ ,
 \\
[\fS^y_m,\fS^z_m]= -i \ {\sf f}(E_m)\frac{\partial \fS^x_m}{ \partial E_m} \ ,
 \\
[\fS^z_m,\fS^x_m]=  -i \ {\sf f}(E_m)\frac{\partial \fS^y_m}{ \partial E_m}\ ,
\end{cases}
\label{ggaudin3}
\end{eqnarray}
which together with Eqs. (\ref{ggaudin}) form an infinite-dimensional
Lie Algebra.

From the Jacobi identities for the generators of this Lie algebra, for
example
\begin{equation}
[\fS^x_n,[\fS^x_m,\fS^y_\ell]] + [\fS^y_\ell,[\fS^x_n,\fS^x_m]] +
[\fS^x_m,[\fS^y_\ell,\fS^x_n]]=0 \ ,
\end{equation}
we obtain, considering the antisymmetry of the functions $X,Y,Z$,
the Gaudin equations \cite{Gau2}
\begin{equation}
Z_{m\ell}X_{\ell n}+Z_{nm}Y_{\ell n}+X_{nm}Y_{m\ell}=0 \ .
\label{gaudineq1}
\end{equation}
Moreover, the relations (for any pair of indices $m$, $\ell$)
\begin{equation}
X_{m\ell}^2-Z_{m\ell}^2=\Gamma_1 \ , \ X_{m\ell}^2-Y_{m\ell}^2=\Gamma_2
\label{gaudineq1p}
\end{equation}
also result from these identities, where $\Gamma_{1,2}$ are constants
independent of $E_m$ and $E_\ell$.

\subsection{Quantum Invariants}
Let us introduce the generalized Gaudin field operators 
\begin{eqnarray}
H(E_m)\equiv H_m=\fS^x_m\fS^x_m +
\fS^y_m\fS^y_m+\fS^z_m\fS^z_m=\frac{1}{2}(\fS^+_m\fS^-_m +
\fS^-_m\fS^+_m+2 \ \fS^z_m\fS^z_m) \ ,
\label{fieldoperator}
\end{eqnarray}
which act on a carrier space ${\cal H}$. These operators are not
the Casimir operators of the GGA since they do not commute with
its generators ($E_m \neq E_\ell$)
\begin{eqnarray}
\begin{cases}
[H_m,\fS^\pm_\ell] = \pm \left ( V^+_ {m\ell} \
\{\fS^\pm_m,\fS^z_\ell\} + V^-_ {m\ell} \ \{\fS^\mp_m,\fS^z_\ell\} -
Z_{m\ell} \  \{\fS^\pm_\ell,\fS^z_m\} \right )\ , \nonumber \\
{[}H_m,\fS^z_\ell{]} = V^+_ {m\ell} \
(\fS^+_\ell \fS^-_m-\fS^+_m \fS^-_\ell) +  V^-_ {m\ell} \
(\fS^+_\ell \fS^+_m -\fS^-_m \fS^-_\ell) \ ,
\end{cases}
\end{eqnarray}
where $\{\hat{A},\hat{B}\}=\hat{A}\hat{B}+\hat{B}\hat{A}$ is the
anticommutator. A key property is that these field operators form
a commutative family
\begin{eqnarray}
[H_m,H_\ell]=0 \ ,
\end{eqnarray}
therefore, they have a common set of eigenvectors in ${\cal H}$
and consequently can be considered as a generating function for
all conserved quantities of quantum integrable systems which will
be called generalized $XYZ$ Gaudin models.

\section{The $XXZ$ Gaudin models}
\label{sec3}

In the following we will concentrate on the diagonalization of the
$XXZ$ Gaudin models, i.e. the cases where $X_{m\ell}=Y_{m\ell}$. As we
will see in the applications, this is the most relevant case from a
physics standpoint, and the simpler mathematically since the
Cartan-Weyl basis is easily defined \cite{future}.


\subsection{Solutions of the $XXZ$ Gaudin equation}
\label{sec3a}

The Gaudin equation Eq. (\ref{gaudineq1}) reduces to
\begin{equation}
Z_{m\ell}X_{\ell n}+Z_{nm}X_{\ell n}+X_{nm}X_{m\ell}=0.
\label{gaudineq2}
\end{equation}
From this expression, together with the antisymmetry of the
functions $X$ and $Z$, one can derive a parametrization for the
coefficients $Z_{\ell n}$ and $X_{\ell n}$:
\begin{eqnarray}
 X_{\ell n} & = & \frac{X_{m\ell}X_{mn}}{Z_{m\ell}-Z_{mn}}
 , \ \
 Z_{\ell n}
    =  \frac{Z_{mn} Z_{m\ell} + X_{mn}^2- Z_{mn}^2}{Z_{m\ell}-Z_{mn}}.
\end{eqnarray}
From the latter expression, and $ Z_{\ell n} = -Z_{n \ell}$, it follows
that
\begin{equation}
  X_{mn}^2- Z_{mn}^2 =  X_{m \ell}^2- Z_{m \ell}^2 = \Gamma,
\label{xzgamma}
\end{equation}
with $\Gamma$ a constant that is independent of any indices.
Taking $E_r$ as a reference parameter, one can write down
\begin{eqnarray}
 X_{\ell n} & = & \frac{X_{r\ell}X_{rn}}{Z_{r\ell}-Z_{r n}}
 , \ \
 Z_{\ell n} =  \frac{\Gamma + Z_{rn} Z_{r\ell}}{Z_{r\ell}-Z_{rn}}.
\end{eqnarray}
Then, the functions $X_{\ell n}$ and $ Z_{\ell n}$, which satisfy the
Gaudin equations Eq. (\ref{gaudineq2}), can be written in terms of a
limited set of parameters $s$, $g$, and $t_i$ as
\begin{eqnarray}
 X_{\ell n} & = & g \frac{\sqrt{1 + s t_\ell^2} \sqrt{1 + s t_n^2}}{t_{\ell}-t_n}
 , \ \
 Z_{\ell n} =  g \frac{1 + s t_\ell t_n}{t_\ell- t_n} ,
 \label{paramxz}
\end{eqnarray}
with
\begin{eqnarray}
 \Gamma & = &  s g^2
 , \ \
  t_i =  -g/Z_{r i},
\end{eqnarray}
where $g$ is a real number and $|s|=0$ or $1$. Taking the limit $t_\ell
\rightarrow t_n=x$, one finds  the limiting behavior defined in Eq.
(\ref{limiting}) as
\begin{equation}
{\sf f}(x) = {\sf g}(x) = {\sf h}(x) = g(1 +s x^2).
\end{equation}
In all practical cases one can take the square roots in Eq.
(\ref{paramxz}) to be real and positive (normally any phase can be
absorbed in the definition of the generators $\mathsf{S}^{+}$ and
$\mathsf{S}^{-}$). Furthermore, the condition that the resulting
exactly-solvable Hamiltonians should be Hermitian leads in most cases
to the condition that the parameters $\Gamma$ and $t_i$ be real. In
this case the parameter $s$ is either $+1, -1$ or $0$.

This corresponds to the three cases discussed by Gaudin \cite{Gau2}:
\begin{enumerate}
\item
{\em Rational:} $\Gamma=0, s=0$,
\begin{eqnarray}
X(\eta_\ell,\eta_n)&=&Z(\eta_\ell,\eta_n)=g  \frac{1}{\eta_\ell-\eta_n},
\label{Rat}
\end{eqnarray}
with $t_i=\eta_i$,
\item
{\em Trigonometric:} $\Gamma>0, s=+1$,
\begin{eqnarray}
X(\eta_\ell,\eta_n)&=& g \frac{1}{\sin(\eta_\ell-\eta_n)} \ , \
Z(\eta_\ell,\eta_n)= g\cot(\eta_\ell-\eta_n) \ ,
\label{Trig}
\end{eqnarray}
with $t_i = \tan(\eta_i)$,
\item
{\em Hyperbolic:} $\Gamma<0, s=-1$,
\begin{eqnarray}
X(\eta_\ell,\eta_n)&=&g \frac{1}{\sinh(\eta_\ell-\eta_n)} \ , \
Z(\eta_\ell,\eta_n)= g \coth(\eta_\ell-\eta_n) \ ,
\label{Hyp}
\end{eqnarray}
with $t_i = \tanh(\eta_i)$.
\end{enumerate}
Note that for these three parametrizations one finds that
the limiting behavior is given by
\begin{equation}
{\sf f}(x) = {\sf g}(x) = {\sf h}(x) = g,
\end{equation}
and that the rational model corresponds to the limit $\eta_\ell
\rightarrow \eta_n$ of both the trigonometric and the hyperbolic model.

Recently, Richardson has proposed a new  family of solutions,
given by \cite{Richx}:
\begin{eqnarray} 
X(z_\ell,z_n)&=&\frac{\sqrt{1+2 \alpha z_\ell + \beta z_\ell^2}\sqrt{1+2
\alpha z_n + \beta z_n^2}}{z_\ell-z_n} \ , \
Z(z_\ell,z_n)= \frac{1+\alpha (z_\ell+z_n) + \beta z_\ell
z_n}{z_\ell-z_n}.
\end{eqnarray}
Evaluating expression Eq. (\ref{xzgamma}) for this parametrization,
 one finds that $ \Gamma =\beta - \alpha^2$.
Hence depending on the sign of $ \beta - \alpha^2$, one finds that this
solution might be expressed as a reparametrization of the rational,
trigonometric or hyperbolic models.

Another useful parametrization is given by
\begin{eqnarray}
X(\eta_\ell,\eta_n) & = & g \frac{2 c_\ell c_n}{c_\ell^2 - c_n^2},  \  \
Z(\eta_\ell,\eta_n) = g \frac{c_\ell^2+ c_n^2}{c_\ell^2-c_n^2},
 \label{Hypc}
\end{eqnarray}
which can be derived from the hyperbolic parametrization, Eq. (\ref{Hyp}),
by taking $ c_i = e^{\eta_i}.$


\subsection{Diagonalizing the $XXZ$ Gaudin models}
\label{sec3b}

To define the representation (or carrier) space ${\cal H}$ of the
generalized $XXZ$ GGA we introduce the lowest-weight vector
$\ket{0}$, such that,
\begin{equation}
\fS^-_m\ket{0}=0 \ \ , \ \ \fS^z_m\ket{0}=F(E_m)\ket{0} \ \forall E_m \ ,
\end{equation}
with $F(E_m)$ the lowest-weight function. Thus, the carrier space
${\cal H}$ is defined as the linear span of the unnormalized vectors
\begin{eqnarray}
\{ \ket{0}, \fS^+_1\ket{0}, \fS^+_1\fS^+_2\ket{0}, \cdots,
\fS^+_1\fS^+_2\cdots\fS^+_m\ket{0}, \cdots \}\ .
\end{eqnarray}

We want now to diagonalize the Gaudin field operators. Using Eqs.
(\ref{ggaudin3}) it turns out that $\ket{0}$ is an eigenstate of $H_m$
with eigenvalue
\begin{eqnarray}
\omega_0(E_m)= F^2(E_m) - {\sf f}(E_m)
\ \frac{\partial }{\partial E_m} F(E_m) \ .
\end{eqnarray}
To solve the general eigenvalue problem
\begin{equation}
H_m\ket{\Phi}=\omega(E_m)\ket{\Phi} \ ,
\label{eiggaudin}
\end{equation}
we propose the Bethe ansatz ($M \in \Integers^+$)
\begin{equation}
\ket{\Phi}=\prod_{\ell=1}^M \fS^+_\ell \ket{0}=
\fS^+_1\fS^+_2\cdots\fS^+_M\ket{0}\ ,
\end{equation}
and  Eq. (\ref{eiggaudin}) is equivalent to
\begin{equation}
(H_m-\omega_0(E_m))\ket{\Phi}=[H_m,\prod_{\ell=1}^M \fS^+_\ell]\ket{0}
\ .
\end{equation}
Thus, the whole problem reduces to compute the commutator
\begin{equation}
[H_m,\prod_{\ell=1}^M \fS^+_\ell] = \sum_{\ell=1}^M \left (
(\prod_{n=1}^{\ell-1} \fS^+_n) \ [H_m,\fS^+_\ell] \
(\prod_{n=\ell+1}^{M} \fS^+_n)\right )
\end{equation}
whose action upon the state $\ket{0}$ can be written as
\begin{equation}
\left[ H_m,\prod_{\ell =1}^{M}S_{\ell }^{+}\right] \left|
0\right\rangle =\sum_{\ell =1}^{M}\left( \prod_{r \left( \neq \ell
\right) }^{M}S_r^{+}\right) \left[ H_m,S_\ell^{+}\right] \left|
0\right\rangle +\frac{1}{2}\sum_{\ell \neq n =1}^{M}\left( \prod_{r
\left( \neq \ell ,n \right) }^{M}S_r^{+}\right) \left[ \left[
H_m,S_\ell^{+}\right], S_n^{+}\right] \left| 0\right\rangle
\end{equation}
which after some algebraic manipulations reduces to
\begin{equation}
\sum_{\ell=1}^M \left( \Gamma - 2 Z_{m\ell} \
F(E_m)+\!\!\!\sum_{n(\neq \ell)=1}^M\!\!\! Z_{m\ell}Z_{mn} \right)
\ket{\Phi} + 2 \sum_{\ell=1}^M \left( X_{m\ell} \
F(E_\ell)+\!\!\!\sum_{n(\neq \ell)=1}^M\!\!\! X_{m\ell} Z_{n\ell} \right)
\hat{\Psi}^+_{\ell m} \ket{0} \ ,
\end{equation}
with $\hat{\Psi}^+_{\ell m}$ given by
\begin{equation}
\hat{\Psi}^+_{\ell m} =  (\prod_{n=1}^{\ell-1} \fS^+(E_n)) \ \fS^+(E_m) \
(\prod_{n=\ell+1}^{M} \fS^+(E_n)) \ .
\end{equation}
Equating to zero all the coefficients in front of $\hat{\Psi}^+_{\ell
m}$, defines a set of nonlinear coupled equations
\begin{equation}
F(E_\ell)+\!\!\!\sum_{n(\neq \ell)=1}^M\!\!\!
Z_{n\ell} =0 \ , \hspace*{2cm} \ell=1,\cdots,M \ ,
\label{Betheeqn}
\end{equation}
termed Bethe's equations, which determine the set of complex numbers
$\{E_m\}$. Once they are solved, one uses these solutions to write down
the eigenvalues
\begin{equation}
\omega(E_m)=\omega_0(E_m)+ \sum_{\ell=1}^M \left( \Gamma -
2 Z_{m\ell} \ F(E_m)+\!\!\!\sum_{n(\neq \ell)=1}^M\!\!\!
Z_{m\ell}Z_{mn} \right) \ .
\end{equation}

\section{Exactly-Solvable Models Derived from the Gaudin algebra}
\label{sec4}

Thus far, we have not assumed any special form for the generators of
the GGA. Let us consider now a possible
realization in terms of generators of the $\bigoplus_\bl
su(2)=su(2)\oplus su(2)\oplus \cdots \oplus su(2)$ algebra, which
satisfy the relations
\begin{equation}
[S^+_\bi,S^-_\bj]=2\delta_{\bi\bj} S^z_\bj \ , \
[S^z_\bi,S^\pm_\bj]=\pm\delta_{\bi\bj} S^\pm_\bj \ , \label{udos}
\end{equation}
with $(S^+_\bj)^\dagger=S^-_\bj$. The set of indices $\bj$ will be
denoted by the symbol ${\cal T}$, whose cardinal is $N_{\cal T}$.
Defining the following operators in terms of the $\bigoplus_\bl su(2)$
generators,
\begin{eqnarray}
\fS^\pm_m=\sum_{\bj \in {\cal T}}X_{m \bj} S^\pm_\bj \ , \
\fS^z_m=-\frac{1}{2} \one - \sum_{\bj \in {\cal T}} Z_{m\bj}  S^z_\bj \ ,
\end{eqnarray}
one has a possible realization of the generators of the GGA,
Eq. (\ref{ggaudin}) \cite{Note1}. Notice that the
$su(2)$ generators are not constrained to be in any particular
irreducible representation. Similarly, one can realize the latter in
terms of the generators of the Lie algebra $\bigoplus_\bl su(1,1)$,
homomorphic to $\bigoplus_\bl su(2)$, which satisfy
\begin{equation}
[K^+_\bi,K^-_\bj]=-2\delta_{\bi\bj} K^z_\bj \ , \
[K^z_\bi,K^\pm_\bj]=\pm\delta_{\bi\bj} K^\pm_\bj \ ,
\label{s11}
\end{equation}
with $(K^+_\bj)^\dagger=K^-_\bj$, obtaining
\begin{eqnarray}
\fS^+_m=\sum_{\bj \in {\cal T}}X_{m \bj} K^+_\bj \ , \
\fS^-_m=-\sum_{\bj \in {\cal T}}X_{m \bj} K^-_\bj \ , \
\fS^z_m=-\frac{1}{2} \one - \sum_{\bj \in {\cal T}} Z_{m\bj}
K^z_\bj \ .
\end{eqnarray}

For the sake of simplicity, we will proceed with the Gaudin operators
defined from the $\bigoplus_\bl su(2)$ generators.  Extension to
$\bigoplus_\bl su(1,1)$ is straightforward after application of the
non-unitary homomorphic mapping: $\fS^+_m \rightarrow \fS^+_m$,
$\fS^-_m \rightarrow -\fS^-_m$,  $\fS^z_m \rightarrow \fS^z_m$. It is
easy to check that the Gaudin field operators, Eq.
(\ref{fieldoperator}), are given by
\begin{eqnarray}
H_m&=&\sum_{\bi \in {\cal T}} Z_{m\bi} S^z_\bi+
\sum_{\bi,\bj \in {\cal T}} \left ( Z_{m\bi} Z_{m\bj} \ S^z_\bi
S^z_\bj +\frac{X_{m\bi} X_{m\bj}}{2} (S^+_\bi S^-_\bj+S^-_\bi
S^+_\bj)\right )+\frac{1}{4} \ ,
\end{eqnarray}
where we assume that $X_{m\bi}=X(E_m,\eta_\bi)$ and
$Z_{m\bi}=Z(E_m,\eta_\bi)$. Bethe's equations, Eq. (\ref{Betheeqn}),
are given by
\begin{equation}
1+2 \sum_{\bj\in {\cal T}} d_\bj Z_{\ell \bj} + 2 \sum_{n(\neq
\ell)=1}^M\!\!\! Z_{\ell n} =0 \ , \hspace*{2cm} \ell=1,\cdots,M \ ,
\label{Betheeqns}
\end{equation}
where $d_\bj$ is the eigenvalue of  $S^z_\bj(K^z_\bj)$, i.e. $S^z_\bj
(K^z_\bj)\ket{0}=d_\bj \ket{0}$.
In the weakly interacting limit the solutions of these equations are
given by the roots of associated Laguerre polynomials (see Appendix
\ref{appendix1}). This establishes a one-to-one correspondence between
the eigenstates of the non-interacting and the weakly-interacting
models, which proves that the Bethe ansatz covers {\em all} eigenstates
and does not contain any spurious solutions for finite values of $g$.

In the strong interaction limit, $g \rightarrow \pm \infty$, some of
the variables diverge to infinity, where again they can be related
to roots of the associated Laguerre polynomials, scaled with a factor
$g$. However, some of the roots can remain finite. These finite roots
are equivalent to the solutions for the Gaudin spin magnets \cite{Gau2}
and can be related to the elementary excitations of the BCS-model in
the canonical ensemble \cite{Exci}.

From the analytic properties of the $X$ and $Z$ function matrices
($\eta_\bi\neq\eta_\bj$)
\begin{equation}
\oint_{\Gamma_\bi} \frac{d E_m}{2\pi i}
X_{m\bi}X_{m\bj}={\sf f}(\eta_\bi)X_{\bi\bj} \ , \ \oint_{\Gamma_\bi} \frac{d E_m}{2\pi i}
Z_{m\bi}Z_{m\bj}={\sf f}(\eta_\bi)Z_{\bi\bj} \ , \ \oint_{\Gamma_\bi} \frac{d E_m}{2\pi i}
Z_{m\bi}={\sf f}(\eta_\bi) \ ,
\end{equation}
where $\Gamma_\bi$ is a contour in the complex-$E_m$ plane encircling
$\eta_\bi$. In this way, one can write down the constants of motion
$R_\bi=\frac{1}{{\sf f}(\eta_\bi)} \oint_{\Gamma_\bi} \frac{d E_m}{2\pi i}
H_m \ $, $[R_\bi,R_\bj]=0$, as
\begin{eqnarray}
R_\bi&=&S^z_\bi+
2\sum_{\bj \in {\cal T} (\ne \bi)} \left (
\frac{X_{\bi\bj}}{2} (S^+_\bi S^-_\bj+S^-_\bi S^+_\bj)+
Z_{\bi\bj} \ S^z_\bi S^z_\bj\right ) \ , \\
R_\bi&=& K^z_\bi-
2\sum_{\bj \in {\cal T} (\ne \bi)} \left (
\frac{X_{\bi\bj}}{2} (K^+_\bi K^-_\bj+K^-_\bi K^+_\bj)-
Z_{\bi\bj} \ K^z_\bi K^z_\bj\right ) \ ,
\end{eqnarray}
for $\bigoplus_\bl su(2)$ and $\bigoplus_\bl su(1,1)$, respectively,
with eigenvalues $r_\bi=\frac{1}{{\sf f}(\eta_\bi)} \oint_{\Gamma_\bi}
\frac{d E_m}{2\pi i} \omega(E_m)$
\begin{equation}
r_\bi=d_\bi \left ( 1+ 2 \sum_\ell Z_{\bi \ell} + 2 \sum_{\bj\in {\cal T}
(\ne \bi)} d_\bj Z_{\bi\bj} \right ) \ .
\end{equation}

A class of Gaudin model Hamiltonians can be written as $H_{\sf G}=\sum_\bi
\varepsilon_\bi R_\bi$, i.e.
\begin{eqnarray}
H_{\sf G} &=& \sum_\bi \varepsilon_\bi S^z_\bi +
\frac{\tilde{g}}{2N}\sum_{\bi,\bj (\bi\neq\bj)}\left (
\tilde{X}_{\bi\bj} (S^+_\bi S^-_\bj+S^-_\bi
S^+_\bj)+2 \tilde{Z}_{\bi\bj} S^z_\bi S^z_\bj\right ) \ ,  \label{gau1}\\
H_{\sf G} &=& \sum_\bi \varepsilon_\bi K^z_\bi -
\frac{\tilde{g}}{2N}\sum_{\bi,\bj (\bi\neq\bj)}\left (
\tilde{X}_{\bi\bj} (K^+_\bi K^-_\bj+K^-_\bi
K^+_\bj)-2 \tilde{Z}_{\bi\bj} K^z_\bi K^z_\bj\right )  \ ,
\end{eqnarray}
for $\bigoplus_\bl su(2)$ and  $\bigoplus_\bl su(1,1)$, respectively.
In the equations above $\varepsilon_\bi$ is an arbitrary real number,
$\tilde{X}_{\bi\bj}=(\varepsilon_\bi-\varepsilon_\bj)X_{\bi\bj}/g$,
$\tilde{Z}_{\bi\bj}=(\varepsilon_\bi-\varepsilon_\bj)Z_{\bi\bj}/g$ are
real symmetric matrix functions ($X_{\bi\bj}=X(\eta_\bi,\eta_\bj),
Z_{\bi\bj}=Z(\eta_\bi,\eta_\bj)$), and $\tilde{g}=g N$ is a $c$-number
of order of magnitude unity because of thermodynamic stability reasons.
Notice that, since $\varepsilon_\bi$ and $\eta_\bi$ are, in principle,
independent parameters, one may take advantage of this freedom to write
down different kinds of mode-dependent interactions (see section
\ref{sec6b}). Moreover, $\varepsilon_\bi$ and $\eta_\bi$ must be
chosen real for $H_{\sf G}$ to be Hermitian. Clearly, there are other
classes of Gaudin models that involve higher-order combinations of the
integrals of motion $R_\bi$.

Linear combinations of the $R_\bi$'s will be used in the next
sections to derive {\it different} exactly-solvable model Hamiltonians.
Different models result from using different realizations of $su(2)$
(or $su(1,1)$). In the following we will use (canonical) fermion and
boson realizations, though, one could have used many others
\cite{reviewbo}, such as $SU(N)$ spins or hard-core particles, leading
to {\it new} exactly-solvable problems all of them having the same
algebraic root. For example, for $SU(2)$ in the spin $\mathsf{S}=1$
irreducible representation one can write down Eq. (\ref{gau1}) (for the
rational case) in terms of $su(3)$ generators ${\cal S}^{\mu\nu}$
($\mu,\nu=0,1,2$) in the fundamental representation as \cite{reviewbo}
\begin{eqnarray}
H_{\sf G} &=& \sum_\bi \varepsilon_\bi ({\cal S}^{11}_\bi-{\cal
S}^{22}_\bi) + \sum_{\bi,\bj (\bi\neq\bj)}  {J}_{\bi\bj} \ ({\cal
S}^{\mu\nu}_\bi {\cal S}^{\nu\mu}_\bj-{\cal S}^{\mu\nu}_\bi \tilde{\cal
S}^{\nu\mu}_\bj) \ ,
\end{eqnarray}
where ${J}_{\bi\bj}=(\varepsilon_\bi-\varepsilon_\bj)X_{\bi\bj}/2$, and
$\tilde{\cal S}^{\nu\mu}$ are the generators of $su(3)$ in the conjugate
representation.

\section{$\bigoplus_\bl su(2)$ Fermionic Representation Models}
\label{sec5}

\subsection{BCS-like Models}
\label{sec5a}

Not many models in condensed matter physics have attracted that
much attention as the Bardeen-Cooper-Schrieffer (BCS) model of
superconductivity \cite{bcs}, a remarkable phenomenon discovered
in 1911 by Gilles Holst and Kamerlingh Onnes \cite{onnes}, and
which is characterized by vanishing electrical resistance and
perfect diamagnetism \cite{tinkham}. Soon after the introduction
of the BCS model in condensed matter, Bohr, Mottelson and Pines
\cite{BMP} applied the BCS theory to the description of pairing
correlations in finite nuclei. The BCS or pairing Hamiltonian
is given by
\begin{equation}
H_{\mathsf{BCS}}=\sum_{\mathbf{l}} \varepsilon_{\mathbf{l}} \
n_{\mathbf{l} }+ \sum_{\mathbf{l} \sigma
\mathbf{l}^{\prime}\sigma^{\prime}} g_{\mathbf{l}
\mathbf{l}^{\prime}}^{\sigma \sigma^{\prime}}
c^\dagger_{\mathbf{l} \sigma}
c^\dagger_{\overline{\mathbf{l}\sigma}}
c_{\overline{\mathbf{l}^{\prime} \sigma^{\prime}}}
c^{\;}_{\mathbf{l}^{\prime}\sigma^{\prime}} \ . \label{HBCS1}
\end{equation}
The operator $c_{\mathbf{l} \sigma}^{\dagger}\left(
c_{\mathbf{l}\sigma}^{\;}\right)$ creates (destroys) a fermion in the
state $\mathbf{l} \sigma$, where $\sigma$ is the third projection of
the internal spin degree of freedom $\mathsf{S}$, $\mathbf{l}$
refers to all other quantum numbers needed to specify completely the
state, and $n_{\mathbf{l}}= \sum_{\sigma}
c_{\mathbf{l}\sigma}^{\dagger} c_{\mathbf{l}\sigma}^{\;}$ is a number
operator. Though, in principle, the state $\overline{\mathbf{l}\sigma}$
could be an arbitrary conjugate state to $\mathbf{l}\sigma$ (only a
bijective relation between conjugate pairs is required), we will
restrict here to time-reversal  conjugate pairs. Under time-reversal
(effected by an antiunitary operator) the position operator stays
unchanged while the (linear or  angular) momentum or spin operators
change sign. Thus, the time-reversal transformation of single-particle
states is specific to the choice of basis. For example, the eigenstates
of a generic  angular momentum operator $\mathbf{J}$, labeled as
$|jm\rangle$, transform as
\begin{equation}
|\overline{jm}\rangle = (-1)^{j-m} |j-m\rangle .
\label{TR}
\end{equation}
Similarly, the time-reversal transformation of an annihilation operator
in a basis of spin and linear momentum is $c_{\overline{\bk \sigma}} =
(-1)^{\mathsf{S}-\sigma} c_{-\bk-\sigma}$, while  for a basis of spin
and position $c_{\overline{{\bf r} \sigma}} = c_{{\bf
r}\overline{\sigma}} =(-1)^{\mathsf{S}-\sigma} c_{{\bf r}-\sigma}$. For the
sake of clarity, we will assume a position basis in the following such
that the time-reversal operation will be referred exclusively to the
internal spin part of the states, i.e. $c_{\overline{\mathbf{l}
\sigma}} = c_{\mathbf{l}\overline{\sigma}}$.

The pairing Hamiltonian (\ref{HBCS1}) with uniform couplings
$g_{\mathbf{l}\mathbf{l}^{\prime}}^{\sigma \sigma^{\prime}}=g/4$
has been solved exactly in full generality by Richardson in a
series of papers in the sixties \cite{Richardson}. This important
development escaped the attention of the condensed matter and
nuclear physics communities until very recently, when the
Richardson's works were rediscovered in the study of ultrasmall
superconducting grains. In order to regain the exact solution we
will now present a specific representation of the $su(2)$
generators in terms of fermions
\begin{equation}
\tau_{\mathbf{l}}^{+}=\frac{1}{2}\sum_{\sigma}c_{\mathbf{l}\sigma}^{\dagger}
c_{\mathbf{l}\overline{\sigma}}^{\dagger}=\left( \tau_{\mathbf{l}
}^-\right)^{\dagger} \quad,~\tau_{\mathbf{l}}^{z}=\frac{1}{2}
\sum_{\sigma}c_{\mathbf{l}\sigma}^{\dagger}
c_{\mathbf{l}\sigma}^{\;}-\frac{1 }{4}\Omega_{\mathbf{l}} \ ,
\label{repF1}
\end{equation}
where the operator
$\tau_{\mathbf{l}}^{+}$ creates a pair of fermions in
time-reversal states and
$\Omega_{\mathbf{l}}=2\tau_{\mathbf{l}}+1$ is the degeneracy of
the state $\mathbf{l}$ related to the pseudo-spin of the state
$\mathbf{l}$. It can be readily verified that the three operators
$\{\tau^\pm_\bl, \tau^z_\bl\}$ satisfy the $su(2)$ algebra
(\ref{udos}).

The integrability of the BCS Hamiltonian (\ref{HBCS1}) was
recently demonstrated \cite{CRS}. It was shown that
$H_{\mathsf{BCS}}$ can be written as a linear combination of the
integrals of motion of the rational family with
$X_{\bi\bj}=Z_{\bi\bj}=g/(\varepsilon_\bi-\varepsilon_\bj)$
(\ref{Rat})
\begin{equation}
H_{\mathsf{BCS}}=\sum_{\mathbf{l}}\varepsilon_{\mathbf{l}} \
R_{\mathbf{l} }+C=\sum_{\mathbf{l}}\varepsilon_{\mathbf{l}}
(2\tau^{z}_\bl+\frac{1}{2} \Omega_{\mathbf{l}}) +g
\sum_{\mathbf{l}\mathbf{l}^{\prime}}\tau_{\mathbf{l}}^{+}
\tau_{\mathbf{l}^{\prime}}^- \ .
\label{HBCSF}
\end{equation}

The complete set of eigenstates of the pairing Hamiltonian are given by the
product wavefunction
\begin{equation}
\left\vert \Psi\right\rangle =\prod \limits_{m=1}^{M}\mathsf{S}_{m}^{+}
\left\vert \nu\right\rangle \quad,~\mathsf{S}_{m}^{+}=\sum_{\mathbf{l }}
X_{m\mathbf{l}} \ \tau_{\mathbf{l}}^{+}= \sum_{\mathbf{l}}\frac{1}{
E_{m}-2\varepsilon_{\mathbf{l}}}\tau_{\mathbf{l}}^{+} \ ,
\label{AnsaF}
\end{equation}
where $\left\vert \nu\right\rangle \equiv$ $\left\vert
\nu_{1},\nu_{2}\cdots,\nu_{L}\right\rangle ,$ with $L$ the total number
of single particle states, is a state of $\nu$ unpaired fermions
($\nu=\sum_{ \mathbf{l}}\nu_{\mathbf{l}}$) defined by
\begin{equation*}
\tau_{\mathbf{l}}^{-}\left\vert \nu\right\rangle =0\quad,~n_{\mathbf{l}}
\left\vert \nu\right\rangle =\nu_{\mathbf{l}}\left\vert \nu\right\rangle \ .
\end{equation*}
The quantum numbers $\nu_{\mathbf{l}}$ are often referred to as Seniority
quantum numbers in the nuclear physics literature.

The total number of particles is $N=2M+\nu$, with $M$ the number of Cooper
pairs. Each eigenstate (\ref{AnsaF}) is completely defined by a set of $M$
spectral parameters (pair energies) $E_m$ which are a particular solution of
the Richardson's equations
\begin{equation}
1+\frac{g}{2}\sum_{\mathbf{l}=1}^{L}\frac{\Omega_{\mathbf{l}}-2\nu_{\mathbf{l
}}}{2\varepsilon _{\mathbf{l}}-E_m}+2g\sum_{\ell \left( \neq m \right)
=1}^{M}\frac {1}{ E_{m}-E_{\ell}}=0 \ .
\label{Rich}
\end{equation}

The eigenvalues of the BCS Hamiltonian are
\begin{equation}
E=\sum_{\mathbf{l}=1}^{L}\varepsilon_{\mathbf{l}} \ \nu_{\mathbf{l}
}+\sum_{m=1}^{M}E_{m} \ .
\label{eigen}
\end{equation}
One can easily relate the spectra of the repulsive ($g>0$) and
attractive ($ g<0$) cases: If one performs the following canonical
particle-hole transformation
\begin{eqnarray}
\begin{cases}
c^{\dagger}_{\mathbf{l} \sigma} \rightarrow c^{\;}_{\mathbf{l} \sigma} \\
c^{\;}_{\mathbf{l} \sigma} \rightarrow c^{\dagger}_{\mathbf{l} \sigma}
\end{cases}
\ ,
\end{eqnarray}
which is not a symmetry (although the interaction term is invariant), $H_{
\mathsf{BCS}}(g)$ transforms as
\begin{eqnarray}
H_{\mathsf{BCS}}(g) \rightarrow \sum_\bl \Omega_\bl \varepsilon_\bl - H_{
\mathsf{BCS}}(-g) \ ,
\end{eqnarray}
indicating the relation between the two spectra.

In recent years, the exact solution of the BCS Hamiltonian has
been recovered in the study of ultrasmall superconducting (for a
review see \cite{Delft}). The specific Hamiltonian for grains
assumes a set of $ L $ equally spaced doubly-degenerate single
particle states. Implying that $ \Omega_{\mathbf{l}}=2$ and
$\varepsilon_{\mathbf{l}}=\mathbf{l}$, with $\mathbf{l}=1,2,\cdots,L$.
The Richardson's equations (\ref{Rich})
reduce to
\begin{equation}
1+g\sum_{\mathbf{l}=1}^{L}\frac{1-\nu_{\mathbf{l}}}{2\varepsilon_{l}-E_m}
+2g\sum _{\ell\left( \neq m \right) =1}^{M}\frac{1}{E_{m}-E_{\ell}}=0 \ ,
\label{Rich1}
\end{equation}
with $\nu_{\mathbf{l}}=0,1$. The ground state for an even number of
particles $N$ is in the sector of no broken pairs, $\nu_{\mathbf{l}}=0$ for
all $\mathbf{l}$, while for odd $N$, $\nu_{\mathbf{l}}=1$ for $\mathbf{l}
=\left( N+1\right) /2$ and zero otherwise. In other words, the Fermi level
is blocked by a single particle, excluding it from the active space as can
be seen from the second term in equation (\ref{Rich1}). The additional gap
at the Fermi energy due to the blocking of this level is at the origin of
the odd-even difference observed in the tunneling spectra of small grains.
The excited states of the model are either collective states (pairing
vibrations) within the same Seniority subspace \cite{Exci}, or
non-collective broken pairs \cite{Exci2}.

While this pairing model has found great success describing the physics of
ultrasmall grains, it can be likewise applied to axially deformed nuclei
with non equally spaced single particle levels. The reason that prevented
its use in standard nuclear structure calculations for so many years, was
the lack of an efficient numerical procedure to solve the equations (\ref
{Rich1}) for a large number of non-equally spaced levels. While in the
equally-spaced case the method proposed by Richardson \cite{Rich2} allowed
the treatment of systems with $\sim10^{3}$ particles \cite{Sie}, the
singularities arising in the numerical solutions of the equations with
non-equally space levels are difficult to treat. Recently, it has been
proposed a new numerical procedure to avoid the singularities which seems to
be very promising \cite{Step}, and it might open the scope for applications
to several quantum systems.

The original BCS model for superconductivity \cite{bcs} was
introduced in the context of bulk metallic superconductors. In
this case, electrons (spin-1/2 fermions) are confined in an
arbitrary dimensional box with periodic boundary conditions with
single-particle states of the Bloch type. Pairing occurs in
momentum $\mathbf{k}$-space. For pairing in the singlet $s$ -wave
channel and Cooper pairs with zero momentum $(\mathbf{k} \uparrow,
- \mathbf{k} \downarrow)$ the BCS Hamiltonian can be written as
(${n}_{\mathbf{ k}\sigma}=c^{\dagger}_{\mathbf{k \sigma}}
c^{\;}_{\mathbf{k\sigma}}$ with $ \sigma=\uparrow,\downarrow$)
\begin{eqnarray}
H_{\mathsf{BCS}}&=&\!\!\sum_{\mathbf{k} \sigma} \varepsilon_{\mathbf{k}
\sigma} \ {n}_{\mathbf{k}\sigma} + g \sideset{}{'}\sum_{\mathbf{k k^{\prime}}
} \ c^{\dagger}_{\mathbf{k\uparrow}}c^{\dagger}_{-\mathbf{k\downarrow}}
c^{\;}_{-\mathbf{k^{\prime}\downarrow}}c^{\;}_{\mathbf{k^{\prime}\uparrow}}
\notag \\
&=&\!\!\sum_{\mathbf{k}} [\varepsilon_{\mathbf{k}}({n}_{\mathbf{k}\uparrow}+
{n}_{\mathbf{-k}\downarrow})-g \ {n}_{{\mathbf{k}}\uparrow} {n}_{{-\mathbf{k}
}\downarrow}]+g \sum_{\mathbf{k k^{\prime}}} \ c^{\dagger}_{\mathbf{k\uparrow
}}c^{\dagger}_{-\mathbf{k\downarrow}} c^{\;}_{-\mathbf{k^{\prime}\downarrow}
}c^{\;}_{\mathbf{k^{\prime}\uparrow}} \ ,  \label{BCS}
\end{eqnarray}
where the prime in the first double sum means that the terms
$\mathbf{k}= \mathbf{k}^{\prime}$ are omitted, and
$c_{\mathbf{k}\sigma }^{\dagger}$ creates an electron with
momentum $\mathbf{k}$ and spin $\sigma$. It has been assumed
time-reversal invariance, i.e. $\varepsilon_{\mathbf{k}
\uparrow}=\varepsilon_{-\mathbf{k} \downarrow}=\varepsilon_\bk$.
The relevant $su(2)$ algebra in this case is
\begin{equation}
{\tau}^{+}_{\mathbf{k}} = c^{\dagger}_{{\mathbf{k}}\uparrow} c^{\dagger}_{{-
\mathbf{k}}\downarrow}=(\tau_\bk^-)^\dagger \ , \ {\tau}^{z}_{\mathbf{k}}=
\frac{1}{2} ({n}_{{\mathbf{k}}\uparrow}+{n}_{{-\mathbf{k}}\downarrow}-1) \ ,
\label{set2}
\end{equation}
where $\Omega_{\mathbf{k}}=2$, i.e. the single particle states $\mathbf{k}
\uparrow$ and $-\mathbf{k} \downarrow$ are degenerate. However, $
\varepsilon_{\mathbf{k}}$ may, in principle, differ from $
\varepsilon_{-\mathbf{k}}$. The Hamiltonians of Eqs. (\ref{BCS}) and (\ref
{HBCSF}) are dynamically equivalent. To see this let us rewrite Eq. (\ref{BCS})
in terms of the pseudospin operators $\tau$
\begin{eqnarray}
H_{\mathsf{BCS}}=\sum_{\mathbf{k}} [\varepsilon_{\mathbf{k}}(2 \tau^z_\bk+1)
-g(\tau^z_\bk+2 (\tau^z_\bk)^2)] +g \sum_{\mathbf{k k^{\prime}}} \ \tau^+_{
\mathbf{k}} \tau^-_{\mathbf{k}^{\prime}}\ .  \label{BCSn}
\end{eqnarray}
It can be easily shown that the operators $\sum_\bk \tau^z_\bk$, and $
\sum_\bk (\tau^z_\bk)^2$ are conserved quantities, i.e., commute with $H_{
\mathsf{BCS}}$. Thus, up to an irrelevant global constant,
\begin{eqnarray}
H_{\mathsf{BCS}}=\sum_{\mathbf{k}} \varepsilon_{\mathbf{k}} (2 \tau^z_\bk+1)
+g \sum_{\mathbf{k k^{\prime}}} \ \tau^+_{\mathbf{k}} \tau^-_{\mathbf{k}
^{\prime}}\ ,  \label{BCSnn}
\end{eqnarray}
which is clearly equivalent to Eq. (\ref{HBCSF}). The eigenvalues of the BCS
Hamiltonian, Eq. (\ref{BCS}), are given by Eq. (\ref{eigen}) where the
parameters $E_{m}$ are the solutions of the Richardson's equations, Eq. (\ref
{Rich}), with $\Omega_\bk=2$. One needs to take into account the fact that
for each $\mathbf{k}$ there is a $-\mathbf{k}$ in those sums. Moreover, if
the crystal has space-inversion symmetry $\varepsilon_{\mathbf{k}
}=\varepsilon_{-\mathbf{k}}$. This additional symmetry, which converts each
single-particle level into a four-fold degenerate one, may have dramatic
consequences. For example, the numerical solution for the ground state in
the BCS case is free of singularities due two the fact that the pair
energies $E_m$ come in complex conjugate pairs for any value of the coupling
strength $g$.

In previous work \cite{reviewbo}, a gauge $SU(2)$ symmetry was identified.
The $SU(2)$ symmetry generators are the local operators
\begin{equation}
{S}^{+}_{\mathbf{k}} = c^{\dagger}_{{\mathbf{k}}\uparrow} c^{\;}_{{-\mathbf{k
}}\downarrow}=(S_\bk^-)^\dagger \ , \ {S}^{z}_{\mathbf{k}}=\frac{1}{2} ({n}_{
{\mathbf{k}}\uparrow}-{n}_{{-\mathbf{k}}\downarrow}) \ ,  \label{set1}
\end{equation}
which commute with the pseudo-spins $\tau_\bk$, i.e $\left [ S_\bk^\mu,
\tau_{\mathbf{k}^{\prime}}^\nu \right ] = 0$, for $\mu,\nu=\pm,z$.
This symmetry amounts to the conservation of the charge parity
per mode pair $(\bk \uparrow,-\bk \downarrow)$. Indeed,
this local symmetry is responsible for the Pauli blocking of the (unpair)
singly-occupied states. We would like to emphasize that all the symmetry
analysis applied to Eq. (\ref{BCS}) is also applicable, after proper
rewritting of the symmetry operators, to Eq. (\ref{HBCSF}).

\begin{figure}
\includegraphics[width=4.3in]{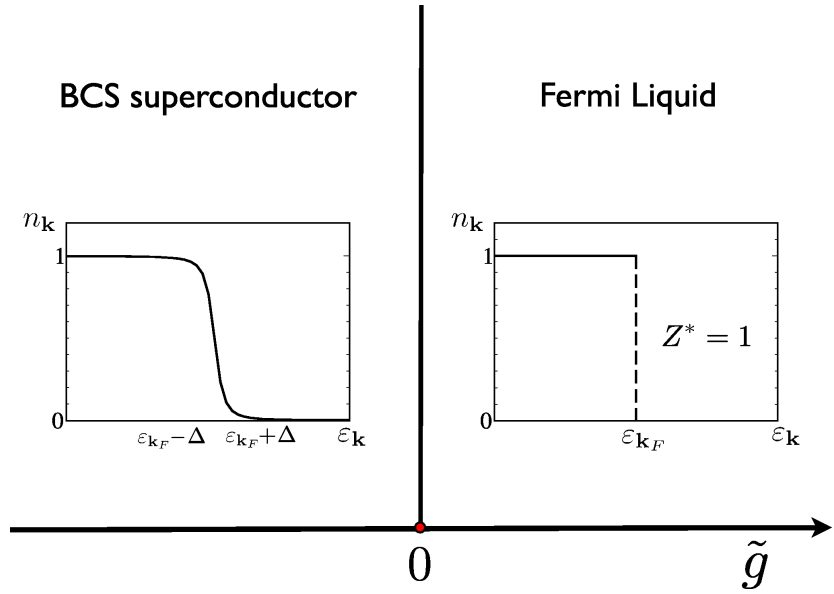}
\caption{Quantum phase diagram of the BCS model as a function of the
interaction strength $\tilde{g}$ in the thermodynamic,  $N \rightarrow
\infty$, limit. The insets represent the single-particle occupation
number $n_\bk$ in each quantum phase. Notice that, for positive
$\tilde{g}$, the Fermi liquid quasiparticle renormalization factor
$Z^*$ is unity regardless of the magnitude of the interaction.}
\label{QPT}
\end{figure}

It is interesting to analyze the quantum phase diagram of the BCS
Hamiltonian $H_{\mathsf{BCS}}$ as a function of the coupling
strength $\tilde{g}=g N$. To this end one needs to study the
behavior of the quantum correlations of the ground state in the
thermodynamic limit. It has been shown, under quite general
assumptions, that the Bethe equations of the integrable BCS
Hamiltonian in the thermodynamic limit are the BCS equations
\cite{Rich, Rom}. The condensation energy for attractive pairing
in this limit is $E_{\sf cond}= -\frac{2 \omega_{D}}{d} e^{2/
\tilde{g}}$ where $d$ is the mean level spacing ($\sim {\sf
Vol}^{-1}$) and $\omega_{D}$ is the Debye frequency cutoff. For
repulsive pairing $E_{\sf cond}=0$. It turns out that there is a
quantum phase transition between a BCS superconductor (broken
$U(1)$ symmetry) and a Fermi liquid of a peculiar type at
$\tilde{g}=0$ (see Fig. \ref{QPT}). It is important to emphasize
that the ground state energy has an essential singularity at
$\tilde{g}=0^-$, implying that it is an infinite-order
(Kosterlitz-Thouless-like) quantum phase transition but with a
broken $U(1)$ symmetry. (Notice that this result is independent of
the space-dimensionality of the problem.) We have numerically
solved the Bethe equations and found that the Fermi liquid has
quasiparticle renormalization factor $Z^*=1$ independently of the
magnitude of $\tilde{g}$; moreover, it displays enhanced
superconducting fluctuations but it is not a superconductor. The
fact that $Z^*=1$ has been previously remarked in \cite{shastry}
using functional integrals.

The exactly solvable Hamiltonian Eq. (\ref{BCS}) can be generalized to the
case of multibands in the following way
\begin{eqnarray}
H_{\mathsf{BCS\mathfrak{n}}}&=&\!\!\sum_{\mathfrak{n}\mathbf{k}}
\varepsilon_{\mathbf{k}}^\fn \ ({n}_{{\mathfrak{n}\mathbf{k}}\uparrow}+ {n}_{
\mathfrak{n}-\mathbf{k}\downarrow}) + \sideset{}{'}\sum_{\mathfrak{n}
\mathbf{k}\mathfrak{n}^{\prime}\mathbf{k}^{\prime}} g_{\mathfrak{n}\mathfrak{
n}^{\prime}} \ c^{\dagger}_{\mathfrak{n}\mathbf{k}\uparrow}c^{\dagger}_{
\mathfrak{n}-\mathbf{k}\downarrow} c^{\;}_{\mathfrak{n}^{\prime}-\mathbf{k}
^{\prime}\downarrow}c^{\;}_{\mathfrak{n}^{\prime}\mathbf{k}
^{\prime}\uparrow} \ ,  \label{twobands}
\end{eqnarray}
where $\mathfrak{n}$ represents the band index, ${n}_{{\mathfrak{n}\mathbf{k}
}\sigma}=c^\dagger_{\mathfrak{n}\mathbf{k}\sigma} c^{\;}_{\mathfrak{n}
\mathbf{k}\sigma}$, and $g_{\mathfrak{n}\mathfrak{n}^{\prime}}=g_{\mathfrak{n
}^{\prime}\mathfrak{n}}$. The prime in the sum means that the terms $(
\mathfrak{n},\mathbf{k})=(\mathfrak{n}^{\prime},\mathbf{k}^{\prime})$ are
excluded. Global symmetries of the model include $\tau^z=\sum_{\mathfrak{n},
\mathbf{k}} \tau_{\mathfrak{n}\mathbf{k}}^z$ and $\sum_{\mathfrak{n},\mathbf{
k}} (\tau_{\mathfrak{n}\mathbf{k}}^z)^2$ with $\tau_{\mathfrak{n}\mathbf{k}
}^z=\frac{1}{2} ({n}_{{\mathfrak{n}\mathbf{k}}\uparrow}+ {n}_{\mathfrak{n}-
\mathbf{k}\downarrow} -1)$. The local $SU(2)$ symmetry has as generators
\begin{eqnarray}
S^{+}_{\mathfrak{n}\mathbf{k}} = c^{\dagger}_{{\mathfrak{n}\mathbf{k}}
\uparrow} c^{\;}_{{\mathfrak{n}-\mathbf{k}}\downarrow}= (S^{-}_{\mathfrak{n}
\mathbf{k}})^\dagger \ , \ S^{z}_{\mathfrak{n}\mathbf{k}}=\frac{1}{2} ({n}_{{
\mathfrak{n}\mathbf{k}}\uparrow}-{n}_{{\mathfrak{n}-\mathbf{k}}\downarrow})
\ .  \label{set3}
\end{eqnarray}
Clearly, the BCS Hamiltonian of Eq. (\ref{BCS}) is a particular case of Eq. (
\ref{twobands}) for a single band ($\mathfrak{n}=1$). It is straightforward
to see that $H_{\mathsf{BCS\mathfrak{n}}}$ is exactly solvable for
interactions

1.- $g_{\mathfrak{n}\mathfrak{n}^{\prime}}=\delta_{\mathfrak{n}\mathfrak{n}%
^{\prime}}g_{\mathfrak{n}}$ (decoupled BCS bands)

2.- $g_{\mathfrak{n}\mathfrak{n}^{\prime}}=g$\ (effective one-band BCS model)

For the particular case of two bands, the Hamiltonian Eq. (\ref{twobands})
might be of interest to describe the phenomenon of two-gap superconductivity
recently observed in materials like MgB$_{2}$. We may consider here an $
SU(2)_{1}\otimes SU(2)_{2}$ structure, with each $SU(2)_{\mathfrak{n}}$
generated by the elements
\begin{equation*}
\tau _{\mathfrak{n}}^{+}=\sum_{\mathbf{k}}c_{\mathfrak{n}\mathbf{k}\uparrow
}^{\dagger}c_{\mathfrak{n}-\mathbf{k}\downarrow }^{\dagger }=\left( \tau _{
\mathfrak{n}}^{-}\right) ^{\dagger }~,~\tau _{\mathfrak{n}}^{z}= \sum_{
\mathbf{k}}\tau^z_{\mathfrak{n}\mathbf{k}} \ .
\end{equation*}
The case of two flat bands $\varepsilon=-\varepsilon_{\mathbf{k}}^1 =
\varepsilon_{\mathbf{k}}^2$ with equal diagonal interaction terms, i.e.
$g_{11}=g_{22}$, can be easily shown to be exactly solvable: By using the
quantum invariants of the $XXZ$ Richardson-Gaudin (RG) models
\begin{eqnarray}
R_\fn = \tau_\fn^z +2\sum_{\fn'\neq\fn} [ \frac{X_{\fn\fn'}}{2}
(\tau_\fn^+ \tau_{\fn'}^-+\tau_\fn^- \tau_{\fn'}^+)+ Z_{\fn\fn'}
\tau_\fn^z \tau_{\fn'}^z] \ ,
\end{eqnarray}
it can be shown that the two-band pairing Hamiltonian is equivalent
(up to an overall constant) to
\begin{equation}
H_{\mathsf{BCS2}}=2\varepsilon \left( \tau
_{2}^{z}-\tau _{1}^{z}\right) +g_{11}\left( \tau _{1}^{+}\tau
_{1}^{-}+\tau _{2}^{+}\tau _{2}^{-}\right) +g_{12}\left( \tau
_{2}^{+}\tau _{1}^{-}+\tau _{1}^{+}\tau _{2}^{-}\right) \ ,
\label{H2Bn}
\end{equation}
where $g_{11}=4\varepsilon Z_{21}$ and $g_{12}=4\varepsilon X_{21}$ are
two arbitrary real numbers (with the parametrization of Eq.
(\ref{paramxz}), $g_{12}^2-g_{11}^2=(4 \varepsilon g)^2 s$).  To
arrive to expression (\ref{H2Bn}) we have used the Casimir invariants
\begin{eqnarray}
\frac{1}{2} (\tau_\fn^+
\tau_{\fn}^-+\tau_\fn^- \tau_{\fn}^+)+ (\tau_\fn^z)^2 ={\sf S(S+1)}
\end{eqnarray}
together with the conservation of $\tau^z=\tau_1^z+\tau_2^z$ and
$(\tau_1^z)^2+(\tau_2^z)^2$.  This is the Hamiltonian originally
proposed by Suhl, Matthias and Walker \cite{SMW} as an extension of the
BCS model, to include situations where the scattering between electrons
from different bands contributes substantially to the resistivity in
the normal state.

The (unnormalized) eigenstates of $H_{\mathsf{BCS2}}$ are given by
(with $t_1=-\eta$ and $t_2=\eta$)
\begin{equation}
\ket{\Psi}= \prod_{\ell=1}^M \left ( \frac{1}{E_\ell+\eta} \
\tau^+_1 + \frac{1}{E_\ell-\eta} \ \tau^+_2 \right ) \ket{\nu} \ ,
\end{equation}
where the spectral parameters $E_\ell$ satisfy Bethe's equations
($d_\pm= d_1 \pm d_2$, and $2d_{1(2)}=\nu_{1(2)}-\Omega_{1(2)}/2$)
\begin{equation}
\varepsilon +  \eta \frac{g_{12} d_+ {E}_\ell -2 \varepsilon g d_-
(1+s E_\ell^2)}{{E}_\ell^2-\varepsilon^2} +2g \varepsilon \sum_{n\left(
\neq \ell\right)=1}^M  \frac{1+s \ {E}_\ell  {E}_n}{{E}_\ell -{E}_n} =0
\ .
\end{equation}
The corresponding eigenvalues can be easily obtained from those of the
integrals of motion.

More complex situations arise in the application of the BCS model
to the spherical nuclear shell model or to finite lattices. As an
example in nuclear physics we will consider the semi-magic Sn
isotopes . These series of nuclei can be modelled by a set of
valence neutrons occupying the single-particle orbits in the
$N=50-82$ shell interacting with a residual BCS Hamiltonian. In
Table \ref{T1} we show the experimental single particle energies
and the corresponding degeneracies in the spherical single
particle basis.

\begin{center}
\bigskip
\begin{table}[htb]
\caption{Single particle energies and degeneracies for the Tin
isotopes in the $N=50-82$ shell.}
\begin{tabular}{|c|c|c|c|c|c|}
\hline
s. p. level & d$_{5/2}$ & g$_{7/2}$ & s$_{1/2}$ & d$_{3/2}$ & h$_{11/2}$ \\
\hline
s. p. energy (Mev) & 0.0 & 0.22 & 1.90 & 2.20 & 2.80 \\ \hline
s. p. degeneracy & 6 & 8 & 2 & 4 & 12 \\ \hline
\end{tabular}
\label{T1}
\end{table}
\end{center}

Richardson's equations for this case have non-equally spaced levels and
variable degeneracies. While the solution to this problem has been found
using standard techniques \cite{electro}, larger systems would require more
sophisticated methods \cite{Step} to avoid the singularities.

A quite similar situation arises in solving BCS Hamiltonian in
finite two dimensional lattices of size $L\times L$ \cite{2D}. The
single particle energies in units of the hopping matrix element
are $\varepsilon_{k}=-2 \left( \cos k_{x}+\cos k_{y}\right) $,
with $k_{\rho}=2\pi n_{\rho}/L$ and $ -L/2\leq n_{\rho}<L/2$.
Table \ref{T2} shows the single particle energies and degeneracies
for a $6\times6$ lattice.

Numerical applications of the RG models to fermionic problems in
nuclear physics and condensed matter have been concentrated on the BCS
Hamiltonian with uniform couplings. The use of the hyperbolic model, a
particular solution of the $XXZ$ generalized Gaudin models, with
non-uniform coupling strength has been suggested in Ref. \cite{ami5}
to describe the physics of multigrain systems, though no practical
applications has been carried out so far.
\begin{center}
\begin{table}[htb]
\caption{Single particle energies and degeneracies for electrons
in a $6\times6$ square lattice.}
\begin{tabular}{|c|c|c|c|c|c|c|c|c|c|}
\hline
s. p. energy & -4 & -3 & -2 & -1 & 0 & 1 & 2 & 3 & 4 \\ \hline
s. p. degeneracy & 2 & 8 & 8 & 8 & 20 & 8 & 8 & 8 & 2 \\ \hline
\end{tabular}
\label{T2}
\end{table}
\end{center}


Another exactly-solvable model with a separable pairing interaction
(SPI) was proposed by Pan, Draayer and Ormand \cite{Pan98}. The
Hamiltonian has degenerate single-particle energies but some structure
in the pairing interaction
\begin{equation}
H_{\sf SPI}= \varepsilon \sum_{\bj} n_{\bj }
       + \sum_{\bj, \bj', m, m'} g_{\bj \bj'}
       a^\dagger_{\bj m} a^\dagger_{\overline{\bj m}}
       a^{\,}_{\overline{\bj' m'}} a_{\bj'm'} \ ,
\label{HSPI}
\end{equation}
with $g_{\bj \bj'} = g c_\bj c_\bj'$. This model can be derived from
the model of Eq. (\ref{gau1}) using the parameterization of Eq.
(\ref{Hypc}) and taking $\varepsilon_\bj = c_\bj^2$. Inserting the
number operator divided by the number of particles and adjusting the
interaction strength, one can cancel out the one-body and two-body
diagonal parts in the Hamiltonian.

Apart from its relevance in the nuclear shell model, the SPI model has
also been used in connection with atomic BECs \cite{Romb02} and for
establishing variational lower bounds on the energy of general two-body
Hamiltonians \cite{Neck01}.


\subsection{Particle-hole-like Models}
\label{sec5b}

It is clear that what is behind the exact solvability of these different
models is a GGA and the existence of certain quantum
invariants. Different representations of the Gaudin operators lead to
different models but all of them with the same dynamics. In this section, we
continue with $su(2)$ fermionic representation models.

In previous section we have written down BCS-like models using an $su(2)$
representation in terms of pseudo-spins $\tau$. We have also seen that there
is another fermionic representation for $su(2)$ in terms of the generators
of Eq. (\ref{set1}). The natural question that arises is: Can we write down
sensible exactly solvable models of interacting fermions in terms of this $
su(2)$ representation? and the simple answer is yes.

The Lipkin-Meshkov-Glick (LMG) model \cite{limegl} was originally
introduced to study phase transitions in finite nuclei. The model
considers $N$ fermions distributed in two $N$-fold degenerate
levels (termed upper and lower shells). The latter are separated
by an energy gap $\varepsilon $
\begin{equation}
H_{\mathsf{LMG}}=\frac{\varepsilon }{2}\sum\limits_{\mathbf{k}\sigma }\sigma
c_{\mathbf{k}\sigma }^{\dagger }c_{\mathbf{k}\sigma }^{\;}+\frac{V}{2N}
\sum\limits_{\mathbf{k}\mathbf{k}^{\prime }\sigma }c_{\mathbf{k}\sigma
}^{\dagger }c_{\mathbf{k}^{\prime }\sigma }^{\dagger }c_{\mathbf{k}^{\prime
}-\sigma }^{\;}c_{\mathbf{k}-\sigma }^{\;}+\frac{W}{2N}\sum\limits_{\mathbf{k
}\mathbf{k}^{\prime }\sigma }c_{\mathbf{k}\sigma }^{\dagger }c_{\mathbf{k}
^{\prime }-\sigma }^{\dagger }c_{\mathbf{k}^{\prime }\sigma }^{\;}c_{\mathbf{
k}-\sigma }^{\;}\ ,  \label{lmghamilt1}
\end{equation}%
with the quantum number $\sigma =\pm $ labelling the level. In Eq. (\ref
{lmghamilt1}) the interaction term $V$ scatters a pair of
particles across the Fermi level, i.e. it is a two particle-hole
interaction, while the term $W$ exchange particles in the two
levels. Upon introducing the collective particle-hole operators
\begin{eqnarray}
\label{pseudospin1} S^+ = \sum\limits_\bk
c_{{\mathbf{k}}+}^{\dagger }c_{{\mathbf{k}}
-}^{\;}=(S^{-})^{\dagger}\ ,\ {S}^{z}= \frac{1}{2}
\sum\limits_{\bk\sigma} \sigma c^\dagger_{\bk \sigma} c^{\;}_{\bk
\sigma} \ ,
\end{eqnarray}
which satisfy the $su(2)$ commutation relations, Eq. (\ref{lmghamilt1})
may be rewritten as
\begin{equation}
\label{Lip0}
H_{\sf LMG}=\varepsilon S^z + \frac{V}{2N} (S^+ S^+ + S^- S^- ) +
\frac{W}{2N} (S^+S^- + S^- S^+) \;.
\end{equation}
As defined by Eq. (\ref{Lip0}), $H_{\sf LMG}$ is invariant under the
inversion symmetry operation $I$ that transforms $(S^x,S^y,S^z)
\mapsto  (-S^x,-S^y,S_z)$, and it also commutes with the (Casimir)
operator $S^2 = (S^+S^-+S^-S^+)/2+(S^z)^2$. Thus, Eq. (\ref{Lip0}) is
equivalent to
\begin{equation}
\label{Lip}
H_{\sf LMG}=\varepsilon S^z + \frac{V}{2N} (S^+ S^+ + S^- S^- ) -
\frac{W}{N} S^zS^z + \frac{W}{N}S^2 \;.
\end{equation}
The Hamiltonian $H_{\sf LMG}$ has a band matrix representation in an
$su(2)$ basis, and it can be easily diagonalized for large values of
$N$. As such, the model has been used as a testing ground for many-body
approximations in nuclear physics. More recently, the simplicity of the
model and the fact that it can be interpreted as a Heisenberg chain
with long range exchange interactions, made it fashionable to study
relations between entanglement and quantum phase transitions.

We will show in section \ref{sec6c2} that the LMG model is exactly
solvable. But before consider the modified problem
\begin{equation}
H_{\mathsf{p-h}}=\sum\limits_{\mathbf{k}\sigma }\frac{\varepsilon _{\mathbf{k
}}}{2}\ \sigma c_{\mathbf{k}\sigma }^{\dagger }c_{\mathbf{k}\sigma }^{\;}+
\frac{W}{2N}\sum\limits_{\mathbf{k}\mathbf{k}^{\prime }\sigma }c_{\mathbf{k}
\sigma }^{\dagger }c_{\mathbf{k}^{\prime }-\sigma }^{\dagger }c_{\mathbf{k}
^{\prime }\sigma }^{\;}c_{\mathbf{k}-\sigma }^{\;}\ ,
\label{lmghamilt2}
\end{equation}
where $\sigma=\pm$ may be now interpreted as a band index. Let us
introduce the following commuting $su(2)$ algebras
\begin{eqnarray}
{S}_{\mathbf{k}}^{+} &=&c_{{\mathbf{k}}+}^{\dagger }c_{{\mathbf{k}}
-}^{\;}=(S_{\mathbf{k}}^{-})^{\dagger }\ ,\ {S}_{\mathbf{k}}^{z}=\frac{1}{2}
({n}_{{\mathbf{k}}+}-{n}_{\mathbf{k}-})\ , \label{set0n}\\
{\tau }_{\mathbf{k}}^{+} &=&c_{{\mathbf{k}}+}^{\dagger }c_{\mathbf{k}
-}^{\dagger }=(\tau _{\mathbf{k}}^{-})^{\dagger }\ ,\ {\tau }_{\mathbf{k}
}^{z}=\frac{1}{2}({n}_{{\mathbf{k}}+}+{n}_{{\mathbf{k}}-}-1)\ ,
\label{set1n}
\end{eqnarray}
in terms of which $H_{\mathsf{p-h}}$ can be written (up to an irrelevant
constant) as
\begin{equation}
H_{\mathsf{p-h}}=\sum\limits_{\mathbf{k}}\varepsilon _{\mathbf{k}}\ S_{
\mathbf{k}}^{z}+\frac{W}{N}\sum\limits_{\mathbf{k}\mathbf{k}^{\prime }}S_{
\mathbf{k}}^{+}S_{\mathbf{k}^{\prime }}^{-}\ .
\label{lmghamilt3}
\end{equation}
$H_{\mathsf{p-h}}$ is dynamically equivalent to $H_{\mathsf{BCS}}$ and,
thus, it is also exactly solvable.

\bigskip

\section{$\bigoplus_{\bl}su(1,1)$ Bosonic Representation Models}
\label{sec6}

\subsection{Bosonic BCS-like Models}
\label{sec6a}

The boson BCS or pairing Hamiltonian can be written in complete analogy
as in the fermion case, Eq. (\ref{HBCS1}),\ as
\begin{equation}
H_{\mathsf{BBCS}}=\sum_{\mathbf{l}}\varepsilon _{\mathbf{l}} \
n_{\mathbf{l}} + \frac{g}{4}\sum_{\mathbf{ll}^{\prime}}
b_{\mathbf{l}}^{\dagger} b_{\overline{\mathbf{l}}}^{\dagger}
b_{\overline{\mathbf{l}^{\prime}}}^{\;}b_{\mathbf{l} ^{\prime}}^{\;} \ ,
\label{HBCSB}
\end{equation}
where $b_{\mathbf{l}}^{\dagger}$ ($b_{\mathbf{l}^{\prime }}^{\;}$)
creates (destroys) a boson in the state $\mathbf{l}$, and
$n_{\mathbf{l}}=b_{\mathbf{ l}}^{\dagger}b_{\mathbf{l}}^{\;}$ is
the number operator. For simplicity, we will consider here scalar
bosons, but an arbitrary internal spin and be easily taken into
account as in the case of fermions. The label $ \mathbf{l}$ is a
short-hand notation for a set of quantum numbers; for example, the
states of a $3D$ isotropic harmonic oscillator potential are
labeled by $ \mathbf{l}\equiv (nlm)$, where $n$ is the oscillator
quantum number, $l$  is the orbital angular momentum and $m$ its
third projection. In Eq. (\ref{HBCSB}) $\overline{\mathbf{l}}$
refers to the time-reversed state of $\mathbf{l}$. Following
(\ref{TR}) the time-reversed annihilation boson operator is
$b_{\overline{\mathbf{l}}}=
b_{\overline{nlm}}=(-1)^{l-m}b_{nl-m}$.

In the following we will be concerned with spin scalar bosons, a
possibility that cannot be realized in fermionic models. It is worth
emphasizing, however, that the exact solution for boson systems can
easily incorporate the spin degree of freedom (integer spin), and there
might be important applications for spinor BECs \cite{bige} not
explored so far.

Once again, Richardson \cite{RichB} determined the complete
spectrum of the boson BCS Hamiltonian of Eq. (\ref{HBCSB}). This
work also escaped the attention of the physics community until
very recently, when the model was shown to be quantum integrable
\cite{2D}. Exactly-solvable generalizations of the uniform pairing
Hamiltonian were proposed and subsequently applied to various
finite Bose systems \cite{DukeB1,DukeB2}. In analogy with the
fermionic systems presentation of previous sections, we will first
introduce a specific representation of the $su(1,1)$ generators
\begin{equation}
K_{\mathbf{l}}^{+}=\frac{1}{2}b_{\mathbf{l}}^{\dagger }b_{\overline{
\mathbf{l}}}^{\dagger }=\left( K_{\mathbf{l}}^{-}\right) ^{\dagger
}\quad ,~K_{\mathbf{l}}^{z}=\frac{1}{2} b_{\mathbf{l}}^{\dagger
}b_{\mathbf{l}}^{\;}+\frac{1}{4}\Omega _{\mathbf{l}}\ ,
\label{rep1}
\end{equation}
where the operator $K_{\mathbf{l}}^{+}$ creates a pair of fermions
in time-reversal states and $\Omega_{\mathbf{l}}=2K
_{\mathbf{l}}+1$ is the degeneracy of the state $\mathbf{l}$
related to the pseudo-spin $K_{\mathbf{l}}$. The operators of Eq.
(\ref{rep1}) satisfy the $su(1,1)$ commutation relations


The BCS Hamiltonian $H_{\mathsf{BBCS}}$ can be derived from the
rational family (\ref{Rat})  as a linear combination of the
integrals of motion
\begin{equation}
H_{\mathsf{BBCS}}=\sum_{\mathbf{l}}\varepsilon _{\mathbf{l}}\ R_{\mathbf{l}%
}\left( \varepsilon \right) +C=\sum_{\mathbf{l}}\varepsilon _{\mathbf{l}%
}(2K _{\mathbf{l}}^{z}-\frac{1}{2}\Omega _{\mathbf{l}})+g\sum_{\mathbf{l%
}\mathbf{l}^{\prime }}K _{\mathbf{l}}^{+}K _{\mathbf{l}^{\prime
}}^{-}\ .
\label{HBCS}
\end{equation}

The complete set of eigenstates of this model are given by the
product wavefunction
\begin{equation}
\left\vert \Psi \right\rangle =\prod\limits_{m=1}^{M}\mathsf{S}_{m}^{+}
\left\vert \nu \right\rangle \quad,~\mathsf{S}_{m}^{+}=\sum_{\mathbf{l}}
X_{m\mathbf{l}}\ K_{\mathbf{l}}^{+} = \sum_{\mathbf{l}} \frac{1}{2
\varepsilon_{\mathbf{l}}-E_{m}}K _{\mathbf{l}}^{+}\ ,
\label{AnsaB}
\end{equation}
where $\left\vert \nu \right\rangle \equiv $ $\left\vert \nu _{1},\nu
_{2}\cdots ,\nu _{L}\right\rangle ,$ with $L$ the total number of
single particle states, is a state of $\nu $ unpaired bosons ($\nu
=\sum_{\mathbf{l}}\nu _{\mathbf{l}}$) defined by
\begin{equation*}
K_{\mathbf{l}}^{-}\left\vert \nu \right\rangle =0\quad ,~n_{\mathbf{l}
}\left\vert \nu \right\rangle =\nu _{\mathbf{l}}\left\vert \nu
\right\rangle \ ,
\end{equation*}
$\nu_{\mathbf{l}}$ are referred as the Seniority quantum numbers.

The total number of particles is $N=2M+\nu $, with $M$ the number of
paired bosons. Each eigenstate $\left\vert \Psi \right\rangle$ is
completely defined by a set of $M$ spectral parameters (pair energies)
$E_{m}$ which are a particular solution of the Richardson's equations
\begin{equation}
1+\frac{g}{2}\sum_{\mathbf{l}=1}^{L}\frac{\Omega _{\mathbf{l}}+2
\nu _{\mathbf{l}}}{2\varepsilon _{\mathbf{l}}-E_{m}}-2g\sum_{\ell \left(
\neq m\right) =1}^{M}\frac{1}{E_{m}-E_{\ell }}=0\ ,
\label{RichB}
\end{equation}
and their eigenvalues are given by
\begin{equation}
E=\sum_{\mathbf{l}=1}^{L}\varepsilon _{\mathbf{l}}\ \nu _{\mathbf{l}}
+\sum_{m=1}^{M}E_{m}\ .
\label{eigenB}
\end{equation}

Hamiltonians constructed as general linear combinations of the
integrals of motion $H=\sum_{\mathbf{l}=1}\varepsilon
_{\mathbf{l}}\ R_{\mathbf{l}}(\eta ) $ have eigenvalues
$E=\sum_{\mathbf{l}=1}^{L}\varepsilon _{\mathbf{l}}\ \left(
r_{\mathbf{l}}+\nu _{\mathbf{l}}\right) $ where $r_{\mathbf{l}}$
is the eigenvalue of the integral of motion $R_{\mathbf{l}}$
\cite{DukeB2}.

\subsection{Pairing Hamiltonians for bosons in confining traps}
\label{sec6b}

As an application of the boson rational family, we will consider the
problem of a boson system confined to a harmonic-oscillator trap and
subject to boson pairing interactions \cite{DukeB1}. The pairing
Hamiltonian with uniform couplings cannot describe the physics of a
trapped boson system, for the following reason. Looking back at the
commutators of the pair operators, Eq. (\ref{s11}), we see that they
are proportional to the degeneracy of the state $\mathbf{l}$ that
appears inside the definition of the generator $K^z_{\mathbf{l}}$
(Eq. \ref{rep1}).
Thus, the matrix elements of the pairing Hamiltonian between states
$\mathbf{l}$ and $\mathbf{l}^{\prime }$ will be proportional to
$\sqrt{\Omega _{\mathbf{l}}\Omega _{\mathbf{l}^{\prime }}}$. In a 3$D$
harmonic oscillator with $\mathbf{l}\equiv (nlm)$, the shell degeneracy
is $ \Omega_{\bl}\sim n^{2}$. On the other hand, the single-particle
energies are $\varepsilon_{\bl}=n$. Thus, the net effect would be the
scattering of boson pairs to high-lying levels with greater probability
than to low-lying levels, producing unphysical occupation numbers. This
was precisely the behavior observed in a numerical solution of
Richardson's equations (\ref{RichB}) for a system of $1000$ bosons with
an attractive pairing strength $g$ \cite{DukeB1}.

We can use the freedom we have in choosing the parameters $\eta_\bl$
entering in the definition of the $R_\bl$ operators to obtain a
physically relevant exactly-solvable model. In order to cancel out the
unphysical dependence of the pair-coupling matrix elements on the
degeneracies, we choose the $\eta_\bl$'s so that $\eta_{\bl}=\left(
\varepsilon_{\bl}\right)^{3}$. The Hamiltonian, which is given by
the linear combination of the new $ R_\bl$'s is
\begin{equation}
H_{\sf TB}=2\sum_{\bl}\varepsilon_{\bl}R_{\bl}=C+\sum_{\bl}
\overline{\varepsilon}_{\bl} n_{\bl}+\sum_{\bl\neq
\bl^{\prime}}g_{\bl\bl^{\prime}}\left[ K_{\bl}^{+}K_{\bl^{\prime}}^- -
n_{\bl}n_{\bl^{\prime}}\right] ,
\label{hrenorm1}
\end{equation}
where
\begin{equation*}
C=\frac{1}{2}\sum_{\bl}\varepsilon_{\bl}\Omega_{\bl}-\frac{1}{4}\sum_{\bl\neq
\bl^{\prime}}g_{\bl\bl^{\prime}}\Omega_{\bl}\Omega_{\bl^{\prime}} \ , \
\overline{\varepsilon}_{\bl}=\varepsilon_{\bl}-\sum_{\bl^{\prime}\left(
\neq \bl\right) }g_{\bl\bl^{\prime}}\Omega_{\bl^{\prime}} \ ,
\end{equation*}
\begin{equation}
g_{\bl\bl^{\prime}}=\frac{g}{2} \ \frac{1}{\varepsilon_{\bl}^{2}+
\varepsilon_{\bl^{\prime}}^{2}+\varepsilon_{\bl} \varepsilon_{\bl^{\prime}}} \ .
\label{hrenorm2}
\end{equation}
The interaction in Eq. (\ref{hrenorm2}) has the nice feature that its
two-body matrix elements decrease with the number of shells, as one would
expect in general. It has the particular property that the interactions of
the pair- and density-fluctuations are strictly the same but opposite in
sign. Taking into account that $\varepsilon_{\bl}$ is proportional to $n$,
the two-body matrix elements in Eq. (\ref{hrenorm1}) cancel out the
dependence on the degeneracies in the effective pair-coupling matrix
elements. Thus, $H_{\sf TB}$ should be more appropriate than $H_{\sf
BBCS}$ when modelling a harmonically-confined boson system with a
pairing-like interaction.

The spectrum of  $H_{\sf TB}$ can be obtained from the eigenvalues
$r_\bl$ of the associated $R_\bl$ operators as $E=2\sum_\bl
\varepsilon_{\bl}~r_\bl$, with the end result being
\begin{equation}
E=\frac{1}{2}\sum_{\bl}\varepsilon_{\bl}\Omega_{\bl}-\frac{1}{4}\sum_{\bl\neq
\bl^{\prime }}g_{\bl\bl^{\prime }}\Omega_{\bl}\Omega_{\bl^{\prime
}}-2g\sum_{\bl p}
\frac{\varepsilon_{\bl}\Omega_{\bl}}{2\varepsilon_{\bl}^{3}-E_{p}} \ .
\label{erenorm}
\end{equation}
(Note that the first two terms of Eq. (\ref{erenorm}) exactly cancel
the constant term $C$ in Eq. (\ref{hrenorm1})).

We solved Richardson's equations for $H_{\sf TB}$  for a system of
$M=500$ boson pairs and $L =50$ oscillator shells. In this case, the
occupation numbers display a reasonable physical pattern, with the
occupancies decreasing monotonically with increasing single-boson
energy \cite{DukeB1}.

In the case of \emph{repulsive} pairing  a highly unexpected feature
was found \cite{DukeB1}. For small values of $g$ the system behaves as
a normal BEC. At a critical value of the pairing strength $g_c$ a
second-order quantum phase transition takes place. The new phase is
characterized by a fragmentation of the condensate with the two lowest
states macroscopically occupied, while the occupation of the other
levels is negligible.

\subsection{Exactly-solvable two-level boson Hamiltonians}
\label{sec6c}

The restriction of the bosonic RG models to two-levels comprises
several well known quantum models. Among them we will discuss the
Interacting Boson Model \cite{IBM}, the LMG model \cite{limegl} and the
two Josephson-coupled BECs Hamiltonian \cite{Jop}. Let us begin by
defining the two integrals of motion from the most general $XXZ$ RG
models
\begin{eqnarray}
R_{a} &=&K _{a}^{z}- X_{12}\left[ K _{a}^{+}K _{b}^{-}+K
_{a}^{-}K _{b}^{+}\right] +2Z_{12} ~K _{a}^{z}K _{b}^{z}
\notag \\
R_{b} &=&K _{b}^{z}+X_{12}\left[ K _{a}^{+}K _{b}^{-}+K
_{a}^{-}K _{b}^{+}\right] -2Z_{12} ~K _{a}^{z}K _{b}^{z} \ ,
\label{R2}
\end{eqnarray}
where the operators $K^\kappa_{\bl}$ are defined in Eqs.
(\ref{rep1}). From Eqs. (\ref{R2}) we observe that the sum gives the
total number of bosons which is a conserved quantity. We are then left
with one independent quantum invariant, that we can take as the
difference between the two integrals of motion to define the
Hamiltonian
\begin{equation}
H_{\sf B2}=\varepsilon \left( R_{b}-R_{a}\right) =\varepsilon \left(
K_{b}^{z}-K _{a}^{z}\right) +2\varepsilon \left [ X_{12} \left( K_{a}^{+}
K _{b}^{-}+K _{a}^{-}K _{b}^{+}\right) -2 Z_{12} K _{a}^{z}K _{b}^{z}\right]  \ .
\label{HRab}
\end{equation}
Using Eq. (\ref{rep1}) we rewrite $H_{\sf B2}$ as
\begin{eqnarray}
H_{\sf B2} &=&\frac{\varepsilon}{2} \left[ (1-Z_{12} \Omega_a) \ n_{b}-
(1+Z_{12}\Omega_b) \ n_{a}\right] +  v \sum_{\alpha,\beta}\left(
b_{\beta}^{\dagger}b_{\overline{\beta}}^{\dagger}
a_{\overline{\alpha}}^{\;}a_{\alpha}^{\;}+a_{\alpha}^{\dagger}
a_{\overline{\alpha}}^{\dagger}b_{\overline{\beta}}^{\;}
b_{\beta}^{\;}\right)+ w \ n_{b}n_{a} + C\ ,
\label{Hd}
\end{eqnarray}
where $v=\frac{\varepsilon }{2}X_{12}$ and $w=-\varepsilon Z_{12}$
are two arbitrary real numbers, and
$b_{\beta}^{\dagger}(a_{\alpha}^{\dagger})$ creates a boson in
level $b(a)$ with an internal quantum number $\beta(\alpha)$. As
usual the bar in the internal labels means a time-reversed state,
and $\Omega_{b(a)}$ is the degeneracy of the level. The constant
term is $C= \frac{\varepsilon}{4}\left( \Omega _{b}-\Omega_{a} -
Z_{12}\Omega_{b}\Omega_{a} \right)$. Using the parametrization of
Eq. (\ref{paramxz}) (with $t_1=-\eta$ and $t_2=\eta$), $4 v^2 -
w^2= s g^2 \varepsilon^2$, which can be positive, negative or zero
depending upon the choice of parameters.

The (unnormalized) eigenstates of $H_{\sf B2}$ are given by
\begin{equation}
\ket{\Psi}= \prod_{\ell=1}^M \left ( \frac{1}{E_\ell+\eta} \
K^+_a + \frac{1}{E_\ell-\eta} \ K^+_b \right ) \ket{\nu} \ ,
\label{StateBos}
\end{equation}
with spectral parameters satisfying Bethe's equations
\begin{equation}
\varepsilon - 2 \eta \frac{4 v d_+ {E}_\ell + \varepsilon g d_-
(1+s E_\ell^2)}{{E}_\ell^2-\varepsilon^2} +2g \varepsilon \sum_{n\left(
\neq \ell\right)=1}^M  \frac{1+s \ {E}_\ell  {E}_n}{{E}_\ell -{E}_n} =0
\ ,
\label{RichBos}
\end{equation}
where $d_\pm= d_a \pm d_b$, and $2d_{a(b)}=\nu_{a(b)}+\Omega_{a(b)}/2$.
The corresponding eigenvalues can be constructed from the integrals of
motion eigenvalues as
\begin{equation}
E_{\sf B2}=4d_a d_b \ w - \varepsilon d_- - 2 \eta
\sum_{\ell=1}^{M}\frac{4 v d_- {E}_\ell + \varepsilon  g d_+ (1+s
E_\ell^2)}{{E}_\ell^2-\varepsilon^2} \ .
\label{EigBos}
\end{equation}

We will discuss next the application of the two-level RG bosonic models to
three well-known quantum models.

\subsubsection{The Interacting Boson Model}
\label{sec6c1}

The Interacting Boson Model (IBM) has been a highly successful
phenomenological model to describe the collective properties of medium
and heavy nuclei. The IBM captures the collective dynamics of nuclear
systems by representing correlated pairs of nucleons with angular
momentum $\hat{L}$ by ideal bosons with the same angular momentum. In
its simplest version, known as IBM1, there is no distinction between
protons and neutrons and only angular momentum $\hat{L}=0(s)$ and
$\hat{L}=2(d)$ bosons are retained. The model has a $U(6)$ group
structure and three possible dynamical symmetry limits representing
well defined nuclear phases: the $U(5)$ symmetry for vibrational
nuclei, the $O(6)$ symmetry for $\gamma $-unstable nuclei, and the
$SU(3)$ symmetry for axially deformed nuclei. In each of the three
limits the Hamiltonian can be expressed in terms of the Casimir
operators of the group decomposition chain. The three limits are then,
exactly solvable with analytic expressions for the eigenstates.

The transition from $U(5)$ to $O(6)$ can be modelled by a boson paring
Hamiltonian for the form
\begin{equation}
H_{\sf IBM1}=x(n_{d}-n_{s})+\frac{1-x}{N}\sum_{\mu =-2}^{2}\left(
d_{\mu}^{\dagger }d_{ \overline{\mu }}^{\dagger }ss+s^{\dagger}
s^{\dagger}d_{\overline{\mu } }d_{\mu }\right) \ ,
\label{IBM1}
\end{equation}
where $N$ is the total number of bosons and $x$ is a parameter that
interpolates between the linear Casimir operator of $U(5)$, for $x=1$,
and the quadratic Casimir operator of $O(6)$, for $x=0$. The
Hamiltonian $H_{\sf IBM1}$ can be derived from Eq. (\ref{Hd}) by making
the following identifications: $d=b$, $ s=a$, $\Omega_{d}=5$,
$\Omega_{s}=1$, $\varepsilon =2x$, $w =0$, and $v=
\frac{\left(1-x\right) }{N}$.

The transition from the spherical vibrational phase ($U(5)$) to the
$\gamma$-unstable deformed phase was studied within the integrable
model described by the Hamiltonian of Eq. (\ref{IBM1}) \cite{DukeB3}.
It was found a second order quantum phase transition for a critical
value of the control parameter $x$. In fact this is a unique point of
second order phase transitions in the complete parameter space of the
most general IBM\ Hamiltonian. The second-order character of the
transition is related to quantum integrability. The ground state
eigenvalue of $H_{\sf IBM1}$  is an analytic function of the control
parameter $x$ and, though level crossings are allowed due to quantum
integrability, there are no level crossings in the low-energy spectrum.

\subsubsection{The Lipkin-Meshkov-Glick Model}
\label{sec6c2}

The LMG model has been extensively used for decades to simulate the
phase transition from spherical to deformed shapes in finite nuclei. As
introduced in section \ref{sec5b}, it is a schematic model describing
the scattering of particle-hole pairs between two shells of different
parity $\sigma$. Though it was known for a long time that the model was
quantum integrable, some analytic solutions were  found only quite
recently \cite{Pan} using the algebraic Bethe ansatz, after having
mapped the model onto a Schwinger-boson representation. Here we will
show that the LMG model is exactly solvable: After a Schwinger-boson
representation of angular momentum operators, we will map the LMG
model onto the two-boson integrable Hamiltonians of Eq. (\ref{Hd}).

In the Schwinger mapping of the $su(2)$ algebra the generators are
expressed in terms of two bosons $a$ and \ $b$ as
\begin{equation}
S^{+}=b^{\dagger }a=\left( S^{-}\right) ^{\dagger }\quad
,~S^{z}=\frac{1}{2} \left( b^{\dagger }b-a^{\dagger}a\right)
=\frac{1}{2}(n_b-n_a) \ ,
\label{Sch}
\end{equation}
with the constraint
\begin{equation}
2 \mathsf{S}=b^{\dagger }b+a^{\dagger }a=n_b+n_a  \ .
\label{Cons}
\end{equation}
Inserting Eq. (\ref{Sch}) into Eq. (\ref{Lip}) we obtain a
Schwinger-boson representation of the LMG Hamiltonian
\begin{equation}
H_{\sf LMG}=\frac{W}{N}\mathsf{S}+\frac{\varepsilon}{2} (
n_b -n_a ) +\frac{V}{2N}\left(
b^{\dagger}b^{\dagger}aa +a^{\dagger }a^{\dagger}bb\right)+
\frac{W}{N} n_b n_a  \ .
\label{HBlip}
\end{equation}

We then recover the two-boson exactly solvable Hamiltonian of Eq.
(\ref{Hd}) with $\Omega_{b}=\Omega _{a}=1$, $v=V/2N$ and $w=W/N$. In
particular, the (unnormalized) eigenvectors and eigenvalues of $H_{\sf
LMG}$ are
\begin{eqnarray}
\ket{\Psi}_{\sf LMG}&=&\prod\limits_{\ell=1}^{M}\left( \frac{
a^{\dagger }a^{\dagger }}{E_{\ell}+\eta}+\frac{b^{\dagger }b^{\dagger
}}{E_{\ell}-\eta} \right) \ket{\nu}  \\
 E_{\sf LMG}&=& E_{\sf B2} - \frac{w}{4} \ ,
\end{eqnarray}
where $E_{\sf B2}$ is given in Eq. (\ref{EigBos}), and the spectral
parameters $E_\ell$ satisfy Eq. (\ref{RichBos}). Notice that, for each
sector $\mathsf{S}$, the number of bosons that enter in the expression
for $\ket{\Psi}_{\sf LMG}$ is constrained to be $2\mathsf{S}=n_a+n_b$.

We would like to emphasize that our expressions are compact forms valid
for any arbitrary set of parameters in the Hamiltonian. In particular,
they are valid for the parameter range  $W^2 < V^2$, which correspond to
the solutions not found in Ref. \cite{Pan}. Since we have shown (using
the weakly-interacting limit solutions) that this Bethe ansatz covers
all possible eigenstates, it implies that the LMG is exactly solvable.

\subsubsection{Two coupled Bose-Einstein Condensates}
\label{sec6c3}

The Josephson effect, predicted more than forty years ago
\cite{josephson}, describes pair tunneling between two superconductors
through an insulating junction. An analogous effect can be realized
with trapped ultracold bosonic gases in two different ways. In the
first setup, two atomic condensates in the same atomic state are
separated by a controllable potential barrier. In the second setup,
atoms are condensed in two overlapping hyperfine states with an
exchange mixing interaction. Both systems are described by the
Hamiltonian \cite{Jop}
\begin{equation}
H_{\sf J}=-\frac{E_{\sf J}}{N}\left( c^{\dagger }d+d^{\dagger }c\right)
+\frac{E_{c} }{4}\left ( c^{\dagger }cc^{\dagger }c+d^{\dagger
}dd^{\dagger }d\right) \ ,
\label{HJ}
\end{equation}
where $c^{\dagger }$and $d^{\dagger}$ are left or right trap boson
creation operators, or they create bosons in two different hyperfine
states, depending upon the particular setup. $E_{\sf J}$ is the
Josephson coupling exchanging bosons between the two states, and
$E_{c}$ is the charging energy. This Hamiltonian has been recently
exactly solved using the algebraic Bethe ansatz \cite{Links}. In fact,
it is no more than another form of the two-level boson pairing
Hamiltonian of Eq. (\ref{Hd}). It can be easily recast in the
two-level form after performing the unitary canonical transformation
\begin{equation}
c=\frac{1}{\sqrt{2}}\left( a-ib\right)
~,~d=\frac{1}{\sqrt{2}}\left( a+ib\right) \ .
\label{Tran}
\end{equation}
Eliminating irrelevant constant terms, the Josephson Hamiltonian can be
rewritten as
\begin{equation}
H_{\sf J}=\frac{E_{\sf J}}{N} ( n_b-n_a)-\frac{E_{c}}{8} \left (
b^{\dagger}b^{\dagger }aa + a^{\dagger }a^{\dagger }bb- 2 n_b n_a \right ) \ ,
\label{HJ2}
\end{equation}
which can be easily related to the LMG Hamiltonian, $H_{\sf LMG}$, by
choosing $\varepsilon=2E_{\sf J}/N$, $V=-NE_c/4=-W$. Therefore, the
physics of the two coupled BECs is completely analogous to that of the
LMG model.

\section{Mixing realizations and representations of the Gaudin field operators}
\label{sec7}

We have already mentioned that our scheme for generating completely
integrable models relies on finding different representations of the
GGA. So far, we have simply concentrated on exactly solvable models
where every single $su(2)$ or $su(1,1)$ generator labeled by an index
in the set $\cal T$ is equivalently represented. We still have the
freedom to mix the representations of these generators and develop new
exactly-solvable model Hamiltonians. The fact that the two $su(2)$
realizations of Eqs. (\ref{set0n}) and (\ref{set1n}) are mutually
commuting was also noted in \cite{ami5}. They called them the spin and
the charge realizations, respectively. This property implies that the
elements of the spin and charge $su(2)$ algebras act on orthogonal
Hilbert spaces, allowing to define an integrable RG model in each space
separately.  Thus, the following Hamiltonian \cite{ami5} was proposed
to study the interplay between pairing correlations and spin-exchange
interactions
\begin{eqnarray}
H_{\sf ch-S} = \sum_\bi \varepsilon_\bi \tau^z_\bi +
\frac{\tilde{g}}{2N}\sum_{\bi,\bj (\bi\neq\bj)}\left (
\tilde{X}_{\bi\bj} (\tau^+_\bi \tau^-_\bj+\tau^-_\bi
\tau^+_\bj)+2\tilde{Z}_{\bi\bj} \tau^z_\bi \tau^z_\bj +
{W}_{\bi\bj} S_\bi \cdot S_\bj \right ) \ , \label{ch-s}
\end{eqnarray}
where we can recognize the most general integrable pairing Hamiltonian,
Eq. (\ref{gau1}), with an additional spin-exchange interaction, which
is also integrable if the matrix ${W}$ is derived from the rational
Gaudin family as ${W}_{\bi\bj}=\tilde{X}_{\bi\bj}'$. The use of the
rational family assures the conservation of the total spin quantum
number as well as the third component of the total spin. The model
Hamiltonian $H_{\sf ch-S}$ can still accommodate a linear term in the
spin variables representing a non-uniform magnetic field, or even more
general $XXZ$ Gaudin models can be implemented in the spin space at the
cost of breaking the spin rotational symmetry.

A numerical study of the interplay between pairing and exchange
interactions in small metallic dots has been carried out in
\cite{Falci} for systems with up to $30$ levels. Even though the model
used was fully integrable, the numerical results were mostly obtained
by large scale diagonalization methods due to the complexity in solving
the two sets of coupled Richardson's equations. The recently developed
numerical techniques \cite{Step} to solve efficiently these set of
nonlinear equations may help to extend these studies to larger grains.

The space orthogonality between the two $su(2)$ fermionic realizations
can also be exploited by defining different integrable models in the
charge and spin sectors. For instance, it would be possible to mix a RG
integrable Hamiltonian in the charge space with a Heisenberg or
Haldane-Shastry model in the spin space. By mixing different
realizations of the GGA, one can generate spin-fermion, spin-boson, or
simply spin models with spins belonging to different irreducible
representations. It turns out that this mixing-representations scheme
might be useful to study decoherence and dynamic phenomena in open
quantum systems where some degrees of freedom correspond to the system
while the others (coupled in a particular way to the system) represent
the thermal bath. In the following we illustrate these ideas starting
with the generalized Dicke (GD) model Hamiltonian.

\subsection{Generalized Dicke models}
\label{sec7a}

The GD models have been recently derived from the $XXZ$ \ RG models by
replacing one of the $su(2)\,$\ copies by a single boson satisfying the
Heisenberg-Weyl algebra \cite{ATMO}. This procedure can be rigorously
followed by expressing the $su(2)$ generators in the Holstein-Primakoff
representation as
\[
S_{0}^{+}=\sqrt{2\mathsf{S}_{0}} \ b^{\dagger}\sqrt{1-\frac{b^{\dagger}
b}{2\mathsf{S}_{0}}}=\left(  S_{0}^{-}\right)^\dagger
\quad,~S_{0}^{z}=-\mathsf{S} _{0}+b^{\dagger}b
\]
where we have distinguished the particular copy of $su(2)$ by the label
$0$, $b^{\dagger}\left(  b\right )$ is the creation (annihilation)
operator of a boson with $\left[  b,b^{\dagger}\right]=1$, and
$\mathsf{S}_{0}$ is the magnitude of the spin of this particular
representation. Since the RG model is integrable for arbitrary spin
values we can then analyze the limit $\mathsf{S}_{0}\rightarrow\infty$,
which implies the replacement
\[
S_{0}^{+}=\sqrt{2\mathsf{S}_{0}} \ b^{\dagger}\text{, }S_{0}^{-}=\sqrt
{2\mathsf{S}_{0}} \ b \ .
\]

We will skip here the derivation of the new class of integrable
spin-boson models that can be found in \cite{ATMO} and present
the final form of the integrals of motion
\begin{eqnarray}
R_{0}&=&\omega
b^{\dagger}b+2\sum_{\bj}\varepsilon_{\bj}S_{\bj}^{z}+V\sum_{\bj}\left(
b^{\dagger}S_{\bj}^{-}+bS_{\bj}^{+}\right) \nonumber \\
R_{\bi}&=&\omega S_{\bi}^{z}+\frac{V^{2}}{2}\sum_{\bj\left(  \neq
\bi\right)} \frac{1}{\varepsilon_{\bi}-\varepsilon_{\bj}}\left[
S_{\bi}^{+}S_{\bj}^{-}+
S_{\bi}^{-}S_{\bj}^{+}+2S_{\bi}^{z}S_{\bj}^{z}\right] -V\left(
b^{\dagger}S_{\bi}^{-}+bS_{\bi}^{+}\right)
-2\varepsilon_{\bi}S_{\bi}^{z} \ .
\label{R0}
\end{eqnarray}

It can be readily verified that the set of operators of Eqs.
(\ref{R0}) are Hermitian, independent, and mutually commuting.
Therefore, they constitute a new class of integrable spin-boson models.
Though they have been derived from the trigonometric family of the RG
models, the set of $L$ operators $R_{\bi}$ are identical to the rational
family of RG models, except for the last two terms of $su(2)$ which are
essential for ensuring the commutation with the new bosonic integral of
motion $R_{0}$.

Any function of these operators defines an integrable Hamiltonian. In
particular, we can recognize $R_{0}$ as a Dicke Hamiltonian describing
the interactions of a multi-atom system with a single-mode radiation
field. Moreover, a linear combination involving the whole set of
integrals of motion, Eqs. (\ref{R0}), gives rise to more general
integrable spin-boson Hamiltonians.

Richardson's ansatz for the common eigenstates of the integrals of
motion is
\begin{equation}
\left\vert \Psi\right\rangle =
{\displaystyle\prod\limits_{\alpha=1}^{M}}
\left(
b^{\dagger}+\sum_{\bi=1}^{L}\frac{V}{y_{\alpha}-\varepsilon_{\bi}}S_{\bi}
^{+}\right)  \left\vert 0\right\rangle
\label{Ansa}
\end{equation}
where $L$ is the total number of $su(2)$ spins and $M$ the total spin
and third component of the system. The $M$ spectral parameters
$y_{\alpha}$ are particular solutions of the set of $M$ of non-linear
coupled Richardson's equations
\begin{equation}
\frac{\omega}{2V}-\frac{1}{2V}y_{\alpha}-\frac{V}{2}\sum_{\bi}\frac
{\mathsf{S}_{\bi}}{2\varepsilon_{\bi}-y_{\alpha}}-V\sum_{\beta\left(  \neq
\alpha\right)  }\frac{1}{y_{\alpha}-y_{\beta}}=0\label{Ric}
\end{equation}

The corresponding expressions for the eigenvalues of Eqs. (\ref{R0}) are
\begin{eqnarray}
r_{0}&=&\sum_{\alpha}y_{\alpha}-\sum_{\bj}\mathsf{S}_{\bj}\varepsilon_{\bj}
\ , \nonumber \\
r_{\bi}&=&-\frac{\mathsf{S}_{\bi}}{2}\left\{
\omega+2\varepsilon_{\bi}-\frac{V^{2}} {2}\sum_{\bj\left(  \neq
\bi\right)  }\frac{\mathsf{S}_{\bj}}{\varepsilon_{\bi}
-\varepsilon_{\bj}}+2V^{2}\sum_{\alpha}\frac{1}{y_{\alpha}-2
\varepsilon_{\bi}}\right\} \nonumber \ .
\end{eqnarray}

\subsection{An exactly-solvable Kondo-like impurity model}
\label{sec7b}

The effect of magnetic impurities in metals has been a subject of
intense debate since the early 1930s. In 1964 J. Kondo \cite{kondo}
made significant progress by providing an explanation to the problem of
the resistance minimum (as a function of temperature) in some metals
such as Au. He recognized that the interaction of a single magnetic
impurity with the conduction electrons is well represented by the
$s$-$d$ exchange Hamiltonian
\begin{equation}
H_{\sf sd}=\sum_{\bk,\bk'} J_{\bk\bk'} (S^-
c^\dagger_{\bk\uparrow}c^{\;}_{\bk'\downarrow}+
S^+c^\dagger_{\bk\downarrow}c^{\;}_{\bk'\uparrow}+S^z
(c^\dagger_{\bk\uparrow}c^{\;}_{\bk'\uparrow}-
c^\dagger_{\bk\downarrow}c^{\;}_{\bk'\downarrow})) \ ,
\end{equation}
where $S^z,S^\pm$ are the spin operators representing the localized
moment of magnitude $\mathsf{S}$, while $c^\dagger_{\bk\sigma}$ creates
a conduction electron with momentum $\bk$ and spin projection $\sigma$.
This type of interaction is characterized by terms in which the spin of
the electron is flipped upon scattering with the impurity and are
essential to understand the logarithmic contribution to the
resistivity, and thus, its minimum. The simplest Hamiltonian
representing the interaction of a localized moment with a band of
itinerant electrons is the Kondo impurity model
\begin{equation}
H_{\sf K}=\sum_{\bk,\sigma} \varepsilon_\bk n_{\bk \sigma}+H_{\sf sd}\ .
\label{kondo}
\end{equation}
It is important to emphasize that the case $\mathsf{S}=1/2$,
$J_{\bk\bk'}=J/N$, and linear (relativistic) dispersion has been
exactly solved by Andrei and Weigmann using the Bethe ansatz
\cite{anwieg}.

To find a new exactly-solvable single impurity Kondo-like model let us
consider the RG constants of motion, where the localized spin of
magnitude $\mathsf{S}$ is singled out
\begin{eqnarray}
R_{0}&=& B S^z+ \sum_{\bj} \left( X_{0\bj} (
S^- S_{\bj}^{+}+S^+ S_{\bj}^{-} ) + 2 Z_{0\bj} S^z S^z_\bj \right)\ ,
\label{kondo1}
\end{eqnarray}
with electron spins given by
\begin{eqnarray}
{S}_{\mathbf{j}}^{+} &=&c_{{\mathbf{j}}\uparrow}^{\dagger
}c_{{\mathbf{j}} \downarrow}^{\;}=(S_{\mathbf{j}}^{-})^{\dagger }\ ,\
{S}_{\mathbf{j}}^{z}=\frac{1}{2}
({n}_{{\mathbf{j}}\uparrow}-{n}_{\mathbf{j}\downarrow})\ .
\label{set0nn}
\end{eqnarray}
As mentioned above the commuting $su(2)$ algebra
\begin{eqnarray}
{\tau }_{\mathbf{j}}^{+} &=&c_{{\mathbf{j}}\uparrow}^{\dagger}
c_{\mathbf{j} \downarrow}^{\dagger }=(\tau _{\mathbf{j}}^{-})^{\dagger}\ ,\
{\tau }_{\mathbf{j}}^{z}=\frac{1}{2}({n}_{{\mathbf{j}}\uparrow}+
{n}_{{\mathbf{j}}\downarrow}-1) \ ,
\label{set1nn}
\end{eqnarray}
is a gauge symmetry of $R_0$, thus one can write down the following
exactly-solvable Gaudin model
\begin{eqnarray}
H_{\sf GI}&=&2\sum_\bj \varepsilon_\bj {\tau}_{\mathbf{j}}^{z} + R_0
\nonumber \\
&=&\sum_\bj \varepsilon_\bj ({n}_{{\mathbf{j}}\uparrow}+
{n}_{{\mathbf{j}}\downarrow}-1) + B S^z
 + \sum_{\bk,\bk'} \left( J_{\bk\bk'}^\perp (S^-
c^\dagger_{\bk\uparrow}c^{\;}_{\bk'\downarrow}+
S^+c^\dagger_{\bk\downarrow}c^{\;}_{\bk'\uparrow})+J_{\bk\bk'}^{\|} S^z
(c^\dagger_{\bk\uparrow}c^{\;}_{\bk'\uparrow}-
c^\dagger_{\bk\downarrow}c^{\;}_{\bk'\downarrow}) \right) \ ,
\end{eqnarray}
where
\begin{equation}
J_{\bk\bk'}^{\perp(\|)} = \frac{1}{N} \sum_\bj  e^{i (\bk -\bk')\cdot
{\bf r}_\bj} X_{0\bj}(Z_{0\bj}) \ , \ \mbox{ and} \ c^\dagger_{\bj\sigma} =
\frac{1}{\sqrt{N}} \sum_\bk e^{i \bk \cdot {\bf r}_\bj} c^\dagger_{\bk\sigma}
\end{equation}
is the Fourier-transformed electron operator.

To derive an exactly-solvable Kondo-like Hamiltonian one considers the
particular case  $J_{\bk\bk'}^{\perp}=J X_{0\bk}\delta_{\bk\bk'}$ and
$J_{\bk\bk'}^{\|}=J Z_{0\bk} \delta_{\bk\bk'}$ in $H_{\sf sd}$, and adds
the conduction band term
\begin{eqnarray}
H_{\sf GK}=\sum_{\bk,\sigma} \varepsilon_\bk {n}_{{\mathbf{k}}\sigma}+B S^z
+ J\sum_{\bk} \left( X_{0\bk} (S^-
c^\dagger_{\bk\uparrow}c^{\;}_{\bk\downarrow}+
S^+c^\dagger_{\bk\downarrow}c^{\;}_{\bk\uparrow})+ Z_{0\bk} S^z
(c^\dagger_{\bk\uparrow}c^{\;}_{\bk\uparrow}-
c^\dagger_{\bk\downarrow}c^{\;}_{\bk\downarrow}) \right) \ ,
\end{eqnarray}
since in this case ${\tau}_{\mathbf{k}}^{z}$ commutes with $R_0$
(Notice that the magnitude of the localized spin is not restricted
to $\mathsf{S}=1/2$). Both, a minimum in the electrical
resistivity, and the formation of a singlet resonance state
characterize the Kondo physics. Clearly, since $H_{\sf GK}$ is
translationally invariant the impurity contribution to the charge
resistivity is zero. However, using this model one may address the
fundamental issue of the formation of the singlet state, writing
down a many-body state that captures the essence of the Kondo
problem.

The (unnormalized) $N$-particles eigenstates of $H_{\sf GK}$ are given by
\begin{equation}
\left\vert \Psi\right\rangle =
{\displaystyle\prod\limits_{\ell=1}^{M}}
\left( X_{\ell0}
S^++\sum_{\bk}X_{\ell \bk} c^\dagger_{\bk\uparrow}
c^{\;}_{\bk\downarrow}\right)  \ket{\sf FS} \ ,
\end{equation}
where $\ket{\sf FS}$ is the tensor product of the state
$\ket{\nu}=\left\vert\nu_{\bk_1} \cdots \nu_{\bk_j} \cdots \right
\rangle$ (of $\nu=\sum_\bk \nu_\bk$ paired ($\nu_\bk=2$) fermions)
with the remaining $N-\nu$ fermions in a ferromagnetic state, and
the lowest-weight spin state $\ket{0}_S$
\begin{equation}
\ket{\sf FS} = \ket{\nu} \otimes \underbrace{\prod_{\bk}
c^\dagger_{\bk\downarrow} \ket{0}}_{N-\nu} \otimes \ket{0}_S \ ,
\end{equation}
while $E_\ell$'s satisfy the Bethe equations
\begin{equation}
\frac{B}{2J}=d_0 Z_{0\ell}+\sum_{\bk} d_\bk Z_{\bk\ell} +\sum_{n(\neq
\ell)=1}^M\!\!\! Z_{n\ell}  \ , \hspace*{2cm} \ell=1,\cdots,M \ ,
\end{equation}
and the energy eigenvalues are given by ($\nu_\bk=0,1,2$)
\begin{equation}
E=\sum_{\bk} \varepsilon_\bk \ \nu_\bk + d_0 \left ( B+ 2J
\sum_\ell Z_{0 \ell}  + 2J \sum_{\bk} d_\bk Z_{0\bk} \right ) \ .
\end{equation}
The case of zero magnetic field ($B=0$) corresponds to the Gaudin
magnet.

Any set of parameters $X_{0 \bk}, Z_{0 \bk}$ for which $ X_{0
\bk}^2-Z_{0 \bk}^2=\Gamma$ (see Eq. (\ref{xzgamma})) leads to an
integrable exactly-solvable model. Particularly, the case of a
spin-isotropic exchange interaction, i.e. $X_{0 \bk}=Z_{0
\bk}=1/(\eta_0-\eta_\bk)$, is of interest since the total spin is a
good quantum number. It is easy to see that if the exchange coupling is
antiferromagnetic (e.g., $J>0$, $\eta_0 > \eta_\bk$, and $B$ smaller
than a critical value) a singlet Kondo many-body state emerges in the
problem. The emergence of this state is formally connected to BCS
superconductivity, and the connection is established through Bethe's
equations. In this way, for $M>1$, we are rigorously connecting the
Kondo {\it resonance} with the Cooper {\it resonance} problems.

%
%

\subsection{Exactly-solvable spin-boson models}
\label{sec7c}

The study of spin-boson systems, i.e. a single spin of magnitude
$\mathsf{S}$ linearly coupled to a thermal bath represented by a set of
harmonic oscillators, is of particular interest in the theory of open
quantum systems. These systems display important features of
decoherence, that is the dynamical loss of quantum coherence because of
the environment. In this section we will describe two interesting
spin-boson  models that are exactly-solvable.

Let us start from the constant of motion $R_0$ of Eq. (\ref{kondo1}) and
add to it the total magnetization symmetry
\begin{eqnarray}
H_{\sf sb1}&=& (B+\omega) S^z+ \sum_{\bj} X_{0\bj} (
S^- S_{\bj}^{+}+S^+ S_{\bj}^{-} ) + 2  \sum_{\bj} Z_{0\bj} S^z S^z_\bj + \omega
\sum_{\bj} S_{\bj}^{z} \ .
\label{spinboson1}
\end{eqnarray}
Following a similar procedure to the one illustrated in section
\ref{sec7a} we represent the $su(2)$ spins $S_\bj$ (in the limit
$\mathsf{S_b} \rightarrow \infty$) as
\[
S_{\bj}^{+}=\sqrt{2\mathsf{S_b}} \ b^{\dagger}_\bj \text{, }S_{\bj}^{-}=\sqrt
{2\mathsf{S_b}} \ b^{\;}_\bj \ , \ S_{\bj}^z=-\mathsf{S_b}+
b^{\dagger}_\bj b^{\;}_\bj \ ,
\]
and choose
\begin{equation}
 X_{0\bj} =  g \left(1+\frac{\varepsilon_\bj^2}{4\mathsf{S_b}} \right) \ , \
 Z_{0\bj} = -g \sqrt{\frac{1}{2\mathsf{S_b}}} \ \varepsilon_\bj \ ,
\label{xz}
\end{equation}
with the resulting Hamiltonian (up to irrelevant constants)
\begin{eqnarray}
H_{\sf sb1}&=& \bar{B} S^z+ \sum_\bj \omega b^{\dagger}_\bj b^{\;}_\bj
+ V \sum_{\bj} (S^- b^\dagger_{\bj}+S^+ b^{\;}_{\bj} )\ ,
\label{spinboson2}
\end{eqnarray}
where $V=g \sqrt{2\mathsf{S_b}}$, and $\bar{B}=\tilde{B}+\omega$, with
$\tilde{B}=B+V \sum_\bj \varepsilon_\bj$. Notice that the localized
spin $S$ may have arbitrary magnitude $\mathsf{S}$. This model
is known to be trivially solvable. Indeed, it is a particular case
of the Fr\"ohlich-like Hamiltonian describing a spin coupled to
longitudinal optical phonons
\begin{eqnarray}
H_{\sf sb1}&=& \bar{B} S^z+ \sum_\bj \omega b^{\dagger}_\bj b^{\;}_\bj
+ \sum_{\bj} (V_\bj^* S^- b^\dagger_{\bj}+V_\bj^{\;} S^+ b^{\;}_{\bj} )\ ,
\label{spinboson2p}
\end{eqnarray}
which is diagonalized by first performing a unitary canonical mapping to
new bosonic modes $a_\bj$
\begin{equation}
\begin{pmatrix}
a_1 \\ a_2 \\ a_3 \\ \vdots \\ a_N
\end{pmatrix}
=
\begin{pmatrix}
v_1    & v_2 & v_3 & \cdots & v_N \\
v_2^*  & A_{22} & A_{23} & \cdots & A_{2N} \\
v_3^*  & A_{32} & A_{33} & \cdots & A_{3N} \\
\vdots & \vdots & \vdots & \cdots & \vdots \\
v_N^*  & A_{N2} & A_{N3} & \cdots & A_{NN} \\
\end{pmatrix}
\begin{pmatrix}
b_1 \\ b_2 \\ b_3 \\ \vdots \\ b_N
\end{pmatrix} \ ,
\end{equation}
where ($\Lambda=\sqrt{\sum_\bj |V_\bj|^2}$)
\begin{equation}
v_\bi= \frac{V_\bi}{\Lambda} \ , \ A_{\bi\bi}=
\frac{|v_{\bi}|^2 -1-v_1^*}{1+v_1} \ , \
A_{\bi\bj}=\frac{v_{\bi}^*v_\bj}{1+v_1} \ ,
\end{equation}
yielding the Hamiltonian
\begin{eqnarray}
H_{\sf sb1}&=& \bar{B} S^z+ \sum_\bj \omega a^{\dagger}_\bj a^{\;}_\bj
+ \Lambda (S^- a^\dagger_{1}+ S^+ a^{\;}_{1} )\ ,
\label{spinboson2pp}
\end{eqnarray}
representing, in the new basis, optical phonons effectively interacting
with a single spin through a single mode.

A more elaborate spin-boson model can be realized using the other
constants of motion. In this way the following model Hamiltonian results
\begin{eqnarray}
H_{\sf sb2}&=& \tilde{B} S^z+ \sum_{\bi, \bj} t_{\bi \bj} (b^\dagger_\bi
b^{\;}_\bj - b^\dagger_\bi b^{\;}_\bi)+ V \sum_{\bj} (S^-
b^\dagger_{\bj}+S^+ b^{\;}_{\bj} )\ ,
\label{spinboson3}
\end{eqnarray}
where $t_{\bi
\bj}=\frac{x_{\bi}-x_{\bj}}{\varepsilon_{\bi}-\varepsilon_{\bj}}$, with
$\varepsilon_{\bi}$ representing the parameters defining the model, Eq.
(\ref{xz}), and $x_{\bi}$ the coefficients in the linear combination of
the integrals of motion.

The eigenstates of $H_{\sf sb2}$ are
\[
\left| \Psi \right\rangle =\prod_{\alpha }\left( y_{\alpha
} \ S^{+}+\sum_{\bj}b_{\bj}^{\dagger }\right) \left| 0\right\rangle
\]
where the spectral parameters $y_{\alpha}$ are the solutions of the
Bethe equations
\[
1-\frac{\Omega _{0}}{2} V \ y_{\alpha }+ V \sum_{\bj}\frac{1+\varepsilon
_{\bj}y_{\alpha }}{y_{\alpha }}-2V\sum_{\beta \left( \neq \alpha \right) }
\frac{y_{\alpha }y_{\beta }}{y_{\beta }-y_{\alpha }}=0 \ ,
\]
and $\Omega_0=2 \mathsf{S} + 1$. The corresponding eigenvalues are
\[
E_{\sf sb2}=\frac{\Omega _{0}}{2}V\sum_{\alpha }y_{\alpha
}-\frac{G}{2}\sum_{j\neq i}\frac{\left( x_{i}-x_{j}\right)
\varepsilon _{i}\varepsilon _{j}}{\varepsilon _{i}-\varepsilon
_{j}} \ .
\]


\section{Differential operator realizations: Schr\"odinger-Gaudin
operators}
\label{sec8}

So far we concentrated on discrete representations which
led to various exactly solvable lattice models. However, it is
well-known that it is possible to realize representations of Lie
algebras in the form of differential operators. As we will see in this
section these representations will lead to models in the continuum.
Basically we will generalize the work pioneered by Ushveridze and
others \cite{ushve}.

Out of the many applications one can foresee we will concentrate on a
single problem. The problem consists of mapping exactly-solvable
lattice models to their equivalent in the continuum. Clearly, in
general, the full spectrum of the lattice will be embedded in the
spectrum of the continuum equivalent.
For the sake of  simplicity we will use the $\bigoplus_\bl su(2)$
representation.

To illustrate the procedure we will use the BCS pairing lattice
model of Eq. (\ref{BCS}) which, after realizing about its underlying
algebraic structure, reads
\begin{equation}
H_{\sf BCS}=2\sum_\bk \varepsilon_\bk \tau_\bk^z+g \sum_{\bk\bk'}
\tau_\bk^+\tau_{\bk'}^- \ .
\label{BCSnew}
\end{equation}
We wonder what the many-body problem in configuration space (a
continuous manifold), to which it maps onto, is. To this end, we will
consider the corresponding $su(2)$ Lie algebra of differential
operators in the tensor product representation
\begin{eqnarray}
\tau_\bk^+&=& z_\bk \ , \nonumber \\
\tau_\bk^-&=&-z_\bk \ \partial^2_{z_\bk}+2\mathsf{S}_\bk \ \partial_{z_\bk}\ , \\
\tau_\bk^z&=&-\mathsf{S}_\bk + z_\bk \ \partial_{z_\bk} \ , \nonumber \\
\end{eqnarray}
with Casimir operator $\tau_\bk^2=\frac{1}{2}
(\tau_\bk^+\tau_\bk^-+\tau_\bk^-\tau_\bk^+)+\tau_\bk^z\tau_\bk^z=
\mathsf{S}_\bk(\mathsf{S}_\bk+1)$ and where it is assumed that the
differential generators act on polynomials in the variables $z_\bk$ of
maximum order $2\mathsf{S}_\bk$. In other words, the order of the
polynomials depends upon the value $\mathsf{S}_\bk$ of the spin
irreducible representation. In this differential operator
representation Eq. (\ref{BCSnew}) can be written
\begin{eqnarray}
H_{\sf BCS}&=&2\sum_\bk \varepsilon_\bk(-\mathsf{S}_\bk + z_\bk \partial_{z_\bk}
) +g \varphi(\{z_\bk\}) \sum_{\bk}  (- z_\bk \partial^2_{z_\bk}+2\mathsf{S}_\bk
\partial_{z_\bk}) \ , \\
H_{\sf BCS}&=&\bar{E}+g \varphi(\{z_\bk\})\sum_{\bk} \frac{1}{4}
[-i\partial_{x_\bk}+A]^2 + V(\{x_\bk\})\ ,
\label{BCScont}
\end{eqnarray}
where $\bar{E}=-2\sum_\bk \varepsilon_\bk \mathsf{S}_\bk$,
$\varphi(\{z_\bk\})=\sum_\bk z_\bk$, $z_\bk=x_\bk^2$,
\begin{equation}
A= 2i \left(\frac{(\mathsf{S}_\bk+\frac{1}{4})}{x_\bk} + \frac{\varepsilon_\bk
x_\bk}{g \varphi(\{z_\bk\})}\right ) \  , \mbox{and } \
4V(\{x_\bk\})= i (\partial_{x_\bk}A) - A^2 \ .
\end{equation}
Notice that, in this language, $H_{\sf BCS}$ represents a
many-particle system in a {\it gauge} field subjected to a potential $V$.

In the case $\mathsf{S}_\bk=1/2$, for all $\bk$'s (no unpair
single-particle states)
\begin{eqnarray}
\ket{\uparrow}_\bk \rightarrow z_\bk \ , \ \ket{\downarrow}_\bk\rightarrow 1
\end{eqnarray}
and the representation space includes polynomials of degree at most 1
in each variable. In the following we will simply concentrate on this
case which corresponds to degeneracy $\Omega_\bk=2$. It is
straightforward to prove that the ansatz wave function
\begin{equation}
\Psi^M(\{s_{\bk}\})=\sideset{}{'}\sum_{\bk_1,\cdots,\bk_M} s_{\bk_1}
s_{\bk_2} \cdots s_{\bk_M}\ ,
\end{equation}
where the prime in the sum means that $\bk_1 \neq \bk_2 \neq \cdots
\neq \bk_M$, and
\begin{equation}
s_{\bk_m} = \frac{z_{\bk_m}}{2\varepsilon_{\bk_m} -E_m} \ ,
\end{equation}
is a solution of Eq. (\ref{BCScont}). Interestingly, the function
$\Psi^M(\{s_{\bk}\})$ represents an elementary symmetric function of
order $M$. It turns out that these functions are the equivalent of the
Richardson's solutions in the continuum, i.e,
\begin{equation}
H\Psi^M(\{s_{\bk}\})=(\bar{E}+\sum_{m=1}^M E_m)\Psi^M(\{s_{\bk}\})\ ,
\end{equation}
with the complex numbers $E_m$ satisfying Richardson's equations
\begin{eqnarray}
1+g\sum_{\mathbf{k}}\frac{1}{2\varepsilon_{\bk}-E_m}
+2g\sum _{\ell\left( \neq m \right) =1}^{M}\frac{1}{E_{m}-E_{\ell}}=0
\ .
\end{eqnarray}

\section{Conclusions}
\label{sec9}

The benefits of having exact solutions to problems involving strongly
interacting many-particle systems are difficult to overstate. New
exactly(or quasi-exactly)-solvable models are always a unique tool to
better understand physical phenomena characterized by non-linear and
non-perturbative effects. Moreover, exactly-solvable models are
excellent testing grounds for approximations to the many-body problem.
In the present work we have explored a generalized Gaudin algebra,
whose invariants provide the generating function for integrable quantum
Hamiltonians called $XYZ$ Gaudin models. These quantum invariants can
be simultaneously diagonalized using the Bethe ansatz. Different
representations of the generators of the generalized Gaudin algebra
realize many well-known physical Hamiltonians including the
Bardeen-Cooper-Schrieffer, Suhl-Matthias-Walker, Interacting Boson
Model of nuclei, Lipkin-Meshkov-Glick, several BEC models, generalized
Dicke, spin-boson, a new Kondo-like, and many other models not yet
exploited in the literature. In this way, we have identified the
underlying algebraic structure, thus providing a unifying framework. An
advantage of the Bethe ansatz for the Gaudin models is that the
physical interpretation of their eigenfunctions is straightforward. The
built-in correlation physics is so transparent that they could have
well been chosen as {\it exact} variational states.

An important question concerns the differences between mean-field
approximations to the eigenfunctions of the Gaudin models and their
exact solutions in the thermodynamic limit. Here, we briefly discussed
issues related to the quantum critical behavior of these models. In
particular, we analyzed the nature of the transition between the
superconducting and Fermi liquid phases in the BCS model. We concluded
that it is of the Kosterlitz-Thouless type independently of the space
dimensionality of the lattice.

A number of applications have been presented with the intention of
illustrating the variety of physics problems described by microscopic
Hamiltonians which belong to the class of $XXZ$ Gaudin models. We have
shown that the Lipkin-Meshkov-Glick model, widely used in nuclear
physics and, more recently, in connection with Quantum Information
Theory, is exactly-solvable. Our proof seems to complete the work
initiated in Ref. \cite{Pan} to the whole parameter space (including
the sector $W^2<V^2$). The exact solvability of the two-level boson
Hamiltonians given in \ref{sec6c} comprises three important models, the
LMG, the IBM (describing the transition from the $U(5)$ to the $O(6)$
dynamical symmetries), and the two Josephson-coupled BECs. All of them
are, therefore, characterized by the same physics.

We have also shown that the $XYZ$ Gaudin equation, Eq.
(\ref{gaudineq1}),
results from the use of the Jacobi identities for the generators of the
algebra, i.e. it is a property of the algebra.
For the $XXZ$ case, we have derived a family of antisymmetric solutions
which includes the rational, trigonometric, hyperbolic, and Richadson's
solution as special cases. Indeed, we have proved that the latter is a
reparametrization of any of the other three.

Solving efficiently the non-linear Bethe equations, which provide the
necessary spectral parameters, is an important technical issue. In this
regard, knowing the strong- and weak-coupling limits of those equations
help to reduce the complexity of the task. In the strong-interacting
limit this analysis was previously done in Ref. \cite{Yu}. In  Appendix
\ref{appendix1}  we have shown that the weakly-coupled limit solutions
of the Bethe equations are given by the roots of Laguerre polynomials,
by transforming those equations into a generalized Stieltjes equation.
In this way, we proved that the Bethe ansatz encompasses all possible
eigenstates, and does not provide spurious solutions for finite
couplings.


\appendix

\section{The weakly interacting limit}
\label{appendix1}
The Bethe equations, Eq. (\ref{Betheeqns}), can be expressed in terms
of the parametrization of Eq. (\ref{paramxz}) as
\begin{equation}
\frac{1 -2 g s \left(\sum_{\bj\in {\cal T}} d_\bj + M -1 \right)}{1 + s t_\ell^2}
   +2 g  \sum_{\bj\in {\cal T}} \frac{d_\bj }{t_\ell- t_\bj}
  + 2 g \sum_{n(\neq \ell)=1}^M \frac{1 }{t_\ell- t_n}
  = 0  , \hspace*{1cm} \ell=1,\cdots,M .
\label{Betheeqnt}
\end{equation}
In the limit $g \rightarrow 0$ one recovers the non-interacting model,
with the variables $t_\ell$ converging to the parameter values $t_\bj$,
depending on the corresponding distribution of the particles or spins.
Therefore one can make the substitution
\begin{equation}
 t_\ell = t_{\bj_\ell} + g x_\ell,
\end{equation}
where $\bj_\ell$ is the index of the parameter value to which $t_\ell$
converges. Then Eq. (\ref{Betheeqnt}) can be rewritten up to first
order in $g$ as
\begin{equation}
   \frac{ 2   d_{\bj_\ell} } {x_\ell}
 + \sum_{n, \, \bj_n = \bj_\ell , \, n \neq \ell} \frac{2}{x_\ell- x_n}
 = -\frac{1 + 2 g \alpha_{\bj_\ell}}{1 + s   t_{\bj_\ell}^2},
\label{Betheeqnx}
\end{equation}
with $\alpha_{\bj_\ell}$ independent of the variables $x_\ell$:
\begin{equation}
 \alpha_{\bj_\ell} =
     \sum_{\bj\in {\cal T}, \, \bj \neq \bj_\ell}
          d_\bj \frac{1 + s t_{\bj_\ell} t_\bj}{t_{\bj_\ell}- t_\bj}
   + \sum_{n=1, \, \bj_n \neq \bj_\ell }^M
          \frac{1 + s t_{\bj_\ell} t_{\bj_n}} {t_{\bj_\ell}- t_{\bj_n}}
   - s (d_{\bj_\ell} + N_{\bj_\ell}-1),
\end{equation}
and where  $N_{\bj_\ell}$ is the number of variables $t_n$
that cluster around the parameter $t_{\bj_\ell}$.

Eq. (\ref{Betheeqnx}) can be transformed into
a {\em Generalized Stieltjes equation}  \cite{Shastry01},
with the solutions given by:
\begin{equation}
 x_\ell = -\frac{1 + s   t_{\bj_\ell}^2} {1 + 2 g \alpha_{\bj_\ell}} r_l,
\end{equation}
where the $r_l$ are the roots of the associated Laguerre polynomials
$L_{N}^k $, with $k=2 d_{\bj_\ell} -1$ and $N=N_{\bj_\ell}$.
Note that the resulting values for the variables $t_\ell$
are correct up to second order in $g$:
\begin{equation}
  t_\ell = t_{\bj_\ell} + g \frac{ {\sf f}(t_{\bj_\ell}) }{1 +
2g\alpha_{\bj_\ell} } r_l \ .
\end{equation}
The variables $r_l$ will be real for $d_{\bj_\ell}>0$ (with $su(1,1)$
realizations, typical for bosons), while for  $d_{\bj_\ell}<0$ (with
$su(2)$ realizations, typical for fermions) the variables $x_\ell$ will
come in complex conjugated pairs, except for one real value in case of
an odd number of variables per cluster.

Because a polynomial of order $N$ has a unique set of $N$ roots,
the weakly interacting limit establishes a one-to-one mapping between
the non-interacting solutions (defined by the number of variables clustered
around each parameter $t_\bj$) and the solutions at finite values of $g$.

\begin{acknowledgments}
This work was supported by the US DOE through Contract No.
W-7405-ENG-36, by the Spanish DGI under grant  BFM2003-05316-C02-02 and
by the Fund for Scientific Research - Flanders (Belgium). Fruitful
discussions with C. D. Batista, V.~G.~Guerguiev, I.~Martin, S.~Trugman
and D.~Van~Neck are acknowledged.
\end{acknowledgments}


\begin{thebibliography}{99}
\bibitem{future} The $XYZ$ Gaudin models will be analized somewhere else.
\end{thebibliography}
\end{document}